\documentclass[10pt]{article}
\usepackage{times,newtxmath}
\usepackage{graphicx}
\usepackage[letterpaper,margin=0.8in]{geometry}
\usepackage[
backend=biber,
style=nature,
autocite=superscript
]{biblatex}
\usepackage{bm} 
\usepackage{physics} 
\usepackage{multirow} 
\usepackage{multicol}
\usepackage[labelsep=none]{caption}

\usepackage{hyperref}
\hypersetup{colorlinks,linkcolor={blue},citecolor={blue},urlcolor={blue}} 

\renewenvironment{abstract}
	{\quotation}
	{\endquotation}

\date{}

\makeatletter
\renewcommand{\fnum@figure}{\textbf{Figure \thefigure.} }
\renewcommand{\fnum@table}{\textbf{Table \thetable.} }
\makeatother

\usepackage{url}

\def\scititle{
	Triggered ferroelectricity in HfO$_2$ from hybrid phonons and higher-order dynamical charges
}

\title{\vspace{-1.5cm}\bfseries \boldmath \scititle}

\author{
    Seongjoo Jung, 
	Turan Birol$^{\ast}$
    \and
	\small Department of Chemical Engineering and Materials Science, University of Minnesota, Minneapolis, 55455, USA \and
	\small$^\ast$Corresponding author. Email: tbirol@umn.edu
}

\begin{document} 

\maketitle

\vspace{-0.8cm}
\begin{abstract} \bfseries \boldmath
Ferroelectric HfO$_2$ has emerged as a highly promising material for high-density nonvolatile memory and nanoscale transistor applications. However, the uncertain origin of polarization in HfO$_2$ limits our ability to fully understand and control its ferroelectricity.
Ferroelectricity, the emergence of a spontaneous and switchable polarization in solids, is conventionally understood to be governed by unstable structural modes (phonons), arising either directly from an unstable polar phonon or indirectly through coupling of unstable nonpolar phonons with a polar mode.
While these `proper' and `improper' mechanisms successfully explain  ferroelectricity for most systems, they do not encompass all possible phenomena. 
Here, we present a novel mechanism of `hybrid-triggered' ferroelectricity, where a polar order emerges through trilinear coupling without any structural instabilities. 
Our group theoretical analysis starting from a high-symmetry reference structure shows that this mechanism is realized in intensely-debated ferroelectric HfO$_2$, along with quantitative confirmation from first-principles calculations. 
We also show that dynamical charges in this material are highly unconventional, and a significant contribution to the total polarization arises solely from high-order couplings of nonpolar phonons. 
These findings underline that even simple crystal structures can host surprisingly complicated interplay between different structural orders, elucidate the origin of ferroelectricity and antiferroelectricity in fluorite-related structures, and provide foundational understanding for designing superior ferroelectric materials.
\end{abstract}

\begin{multicols}{2}

\begin{refsection} 

\noindent

\section{Introduction}

The origin of ferroelectricity is typically classified as either \textit{proper}, where polarization arises from an unstable polar structural mode \cite{note_mode} (phonon), or \textit{improper}, where polarization is induced by one or more unstable nonpolar modes through their coupling with a polar mode \cite{toledano1987landau,benedek2022hybrid}. Both mechanisms rely on the existence of a phonon instability. Order parameters associated with unstable modes are termed \textit{primary}, while those emerging through coupling with these modes are referred to as \textit{secondary}. Phenomenological theories of ferroelectrics that focused on mechanisms involving unstable phonons have been widely successful in explaining and predicting materials' properties \cite{van2004origin,fennie2005ferroelectric,bousquet2008improper}. 

However, a system where such an approach has not been as successful is the ferroelectric HfO$_2$-based thin-films \cite{boscke2011ferroelectricity}. 
HfO$_2$ is often dubbed the ``ferroelectric of the future'', and has many properties that make it suitable for applications, including compatibility with existing silicon technology \cite{cheema2020enhanced,lee2020scale,liu2019origin, dutta2021piezoelectricity,salahuddin2008use,alam2019critical}. 
While the ferroelectric phase of HfO$_2$ has been most commonly identified with the orthorhombic space group Pca2$_1$ (\#29), the origin of ferroelectricity in this phase remains elusive. This lack of understanding limits the ability to effectively control unfavorable properties, such as the high coercive field, slow domain wall propagation and formation of non-ferroelectric phases \cite{zhou2015electric,buragohain2018nanoscopic,schroeder2022fundamentals, schimpf2025interface}. 

HfO$_2$ exhibits characteristics that both align with and challenge aspects of both proper and improper mechanisms. The absence of a polar structural instability, specifically the $\Gamma_4^-$ mode, suggests that the ferroelectricity in fluorite HfO$_2$ (Fm$\bar{3}$m, \#225) is improper in nature \cite{lee2020scale,cheema2022emergent}. Several other factors also support this theory. 
While the Born effective charges are anomalous ($+38\%$) they are not as enhanced compared to typical proper ferroelectric oxides \cite{ghosez1998dynamical,zhao2002first,van2004origin,fan2022vibrational}, and depolarization field does not effectively suppress the polarization \cite{sai2009absence,cheema2020enhanced,cheema2022emergent}. 
Also, out-of-plane polarization is induced when an in-plane tensile strain is applied \cite{hyuk2014effects}. Together with the negative piezoelectric response, they suggest strong interactions between polar and nonpolar modes. However, the couplings necessary for conventional improper ferroelectricity are forbidden in fluorite structure \cite{zhou2022strain}. 

This observation led to proposal of proper ferroelectricity in HfO$_2$. Notably, HfO$_2$-based thin-films are not reported to encounter the same challenges in electric switching associated with improper ferroelectrics \cite{oh2015experimental}, and a drastic dielectric enhancement is observed near the ferroelectric transition temperature \cite{schroeder2022temperature}. 
Alternative pathways to ferroelectricity have been suggested, involving lower symmetry parent structures such as Pcam (\#57) \cite{qi2025sign,aramberri2023theoretical}, Pcnb (\#60) \cite{raeliarijaona2023hafnia}, Cmme (\#67) \cite{qi2025competing} or P4$_2$/nmc \cite{cao2024softening} phases.
However, the explanations of proper ferroelectricity in HfO$_2$ requires one or more of the following assumptions, each with certain limitations \cite{zhao2025nature}: a higher energy parent structure, a substantial strain, or polar phases that differ from the experimentally observed Pca2$_1$ phase. Importantly, the proposed mechanisms for proper ferroelectricity do not fundamentally align with the antiferroelectric-ferroelectric continuity \cite{boscke2011ferroelectricity,cheema2022emergent,muller2012ferroelectricity} in fluorite systems. Instead, they predict paraelectricity when ferroelectricity is suppressed, and does not account for change in energy curvature at a polarized state \cite{jung2024octahedral}.

\begin{figure*}[t]
	\centering
	\includegraphics[width=0.75\textwidth]{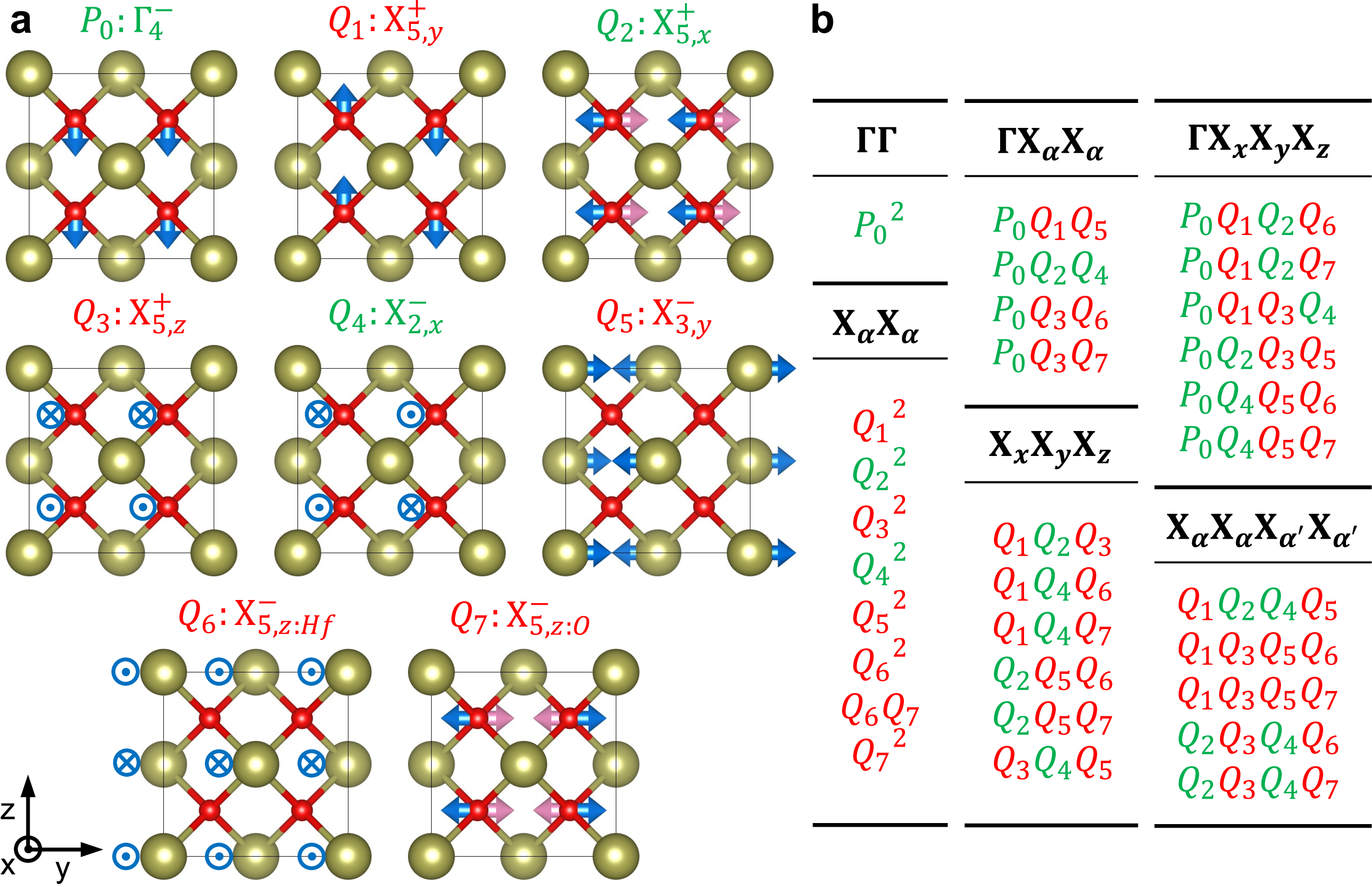} 

	\caption{\textbf{Modes and unique couplings in fluorite HfO$_2$.}
        (a) Eight modes observed in the ferroelectric Pca2$_1$ phase of HfO$_2$. Blue arrows and signs represent ionic displacements in the front half of the unit cell ($x=0.75$ in direct coordinates for oxygen), and the pink arrows represent those in the rear half ($x=0.25$).
        (b) Unique 2nd, 3rd, and 4th-order couplings among modes. Modes that can condense in the dielectric phase are highlighted in green, whereas modes that condense in the triggered phase are shown in red.}
	\label{modes_coupling} 
\end{figure*}

There are other explanations for ferroelectricity in HfO$_2$ that do not fit into the conventional categories of proper or improper ferroelectrics. One notable mechanism is by the negative biquadratic coupling between polar ($\Gamma_4^-$) and nonpolar modes (X$_{5,y}^+$) \cite{holakovsky1973new}. A large tensile strain along one of the in-plane axes is reported to induce the nonpolar mode instability, which in turn, leads to a polar instability through the biquadratic coupling \cite{zhou2022strain}. A trilinear coupling of modes derived from the parent phase P4$_2$/nmc was also identified as a possible origin of ferroelectricity \cite{delodovici2021trilinear}. The polar phase of ZrO$_2$, which shares the same symmetry, has been also investigated and attributed to non-linear interactions between the $X_5^+$ modes, represented by a spline function \cite{reyes2014antiferroelectricity}. 

While these studies offer valuable insight, there is no consensus on a model that explains the ferroelectricity in Pca2$_1$ HfO$_2$. In this study, we employ symmetry-guided expansions of energy and polarization by Landau-Ginzburg-Devonshire (LGD) theory \cite{levanyuk2020landau} with first-principles density functional theory (DFT) involving all modes in Pca2$_1$ HfO$_2$, starting from the highest-symmetry structure. Our results reveal two key, previously unreported findings. First, ferroelectricity in HfO$_2$ is driven by what we dub ``hybrid-triggered'' mechanism involving multiple trilinear and quadlinear couplings of stable modes. Second, the hybrid zone-boundary modes---combinations of multiple nonpolar modes---exhibit exceptionally high polarization, contributing over $-40$\% of the bulk value and offering novel properties from previously unexplored interplay between dipoles.

\section{Results and Discussion}

\subsection{Mode Decomposition and Symmetry-Allowed Couplings}

We begin by analyzing the crystal structure of the polar phase of HfO$_2$ (orthorhombic Pca2$_1$) by decomposing the difference between it and the high-symmetry reference structure (fluorite Fm$\bar{3}$m) into the symmetry-modes. There are eight relevant modes, which are labeled by the irreducible representations (irreps) of the parent space group as shown in \textbf{Figure~\ref{modes_coupling}} \cite{reyes2014antiferroelectricity,note_normal_mode}. Note that the biaxial strain imposes a tetragonal I4/mmm (\#139) parent phase, but we reference irreducible representations from the cubic phase for direct connection with existing literature \cite{reyes2014antiferroelectricity}. Irreducible representations from I4/mmm parent phase and corresponding displacement patterns can be found in Table S1.

$P_0$ denotes the order parameter of the polar zone-center mode ($\Gamma_4^-$), while the remaining parameters, $Q_i$, correspond to nonpolar zone-boundary modes. $Q_4$ represents the only unstable mode ($X_{2,x}^-$) which has non-zero amplitude in the centrosymmetric Aeaa (\#68) phase. The Aeaa phase is a more likely candidate for the nonpolar structure of HfO$_2$ under tensile strain than the frequently assigned P4$_2$/nmc (\#137) phase according to the energies from DFT (Figure S1), consistent with the experimental observations that the long axis of the nonpolar phase has to point towards in-plane direction for the formation of out-of-plane polarization \cite{hyuk2014effects} and changes in interplanar spacing between polar and nonpolar phases \cite{cheema2020enhanced,cheema2022emergent}. The Aeaa phase is also more stable compared to Pcnb (\#60) phase derived from P4$_2$/nmc across the entire range of biaxial strain.

The wavevectors of the zone-boundary modes are essential for understanding the couplings (Figure S2). These wavevectors are not necessarily parallel to the direction of ionic displacements in real space; for instance, the ionic displacements of $Q_1$ are along the $z$ direction, while the wavevector points to the $y$ direction. Unlike previous studies that label the modes by ionic displacement directions \cite{lee2020scale,zhou2022strain,reyes2014antiferroelectricity}, we label them by their wavevectors.

\begin{figure*}[t]
	\centering
	\includegraphics[width=1.00\textwidth]{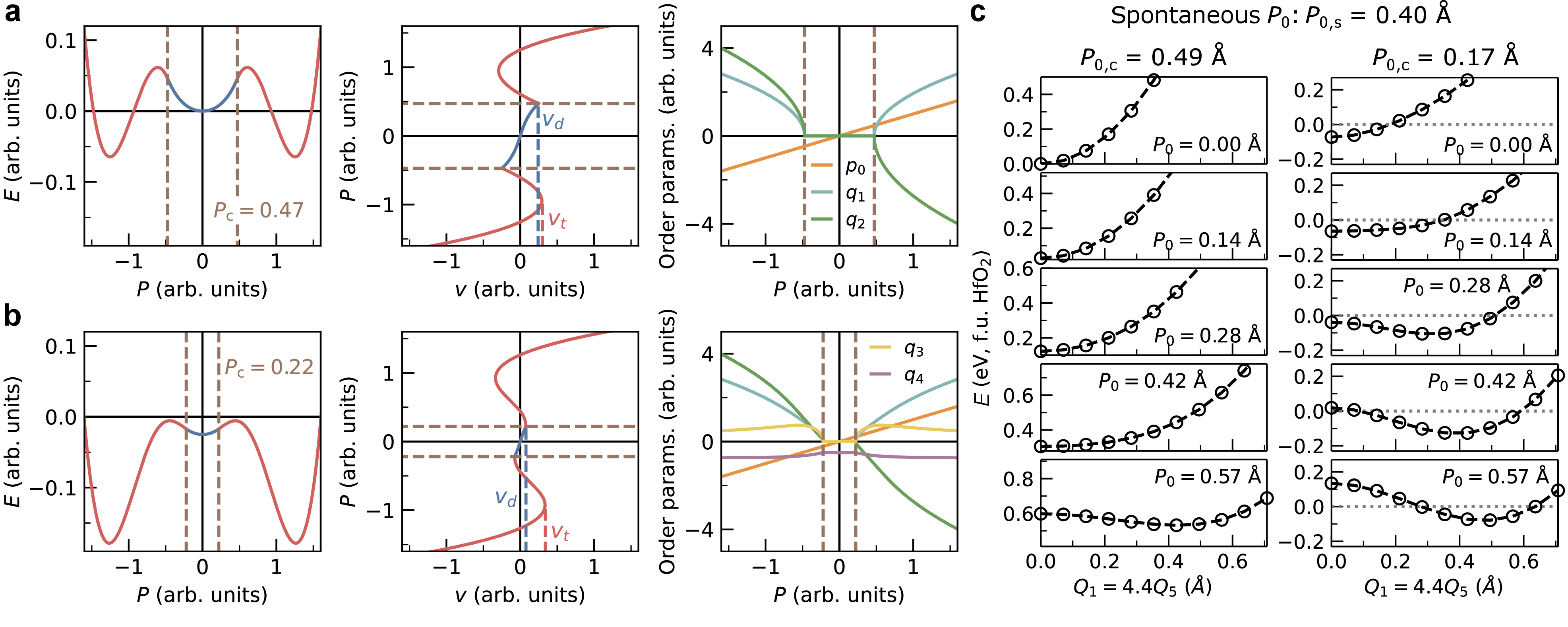} 

	\caption{\textbf{LGD model and DFT validation of hybrid-triggered ferroelectricity.}
    (a) Solution of the theoretical LGD model for hybrid-triggered ferroelectricity described by Equation~(\ref{simple}), formulated in terms of hypothetical order parameters $p_0$, $q_1$, and $q_2$. (b) Same as part (a), but for Equation~(\ref{eq:multicoupling}).  The inclusion of an additional trilinear coupling term, $q_1 q_3 q_4$, significantly reduces the critical polarization required to induce the triggered ferroelectric transition.
    (c) DFT-calculated energies of HfO$_2$ as functions of the structural order parameters $P_0$, $Q_1$, and $Q_5$, providing first-principles validation of hybrid-triggered ferroelectricity. The amplitudes are constrained such that $Q_1/Q_5$ is 4.4 (Figure S4). Results are shown for cases where all other order parameters are constrained to zero (left) and where they are fully relaxed (right), revealing a substantial reduction of the critical polarization $P_{\mathrm{0,c}}$ from 0.49 Å to 0.17 Å.
    }
	\label{multicoupling}
\end{figure*}

Combinations of order parameter components that remain invariant under all symmetry operations of the space group (and hence transform as the $\Gamma_1^+$ irrep) constitute the most general invariant polynomial. Figure~\ref{modes_coupling}b lists the unique mode couplings that are present in this polynomial. Only terms derived from these couplings contribute to the LGD energy in its Taylor expansion around the high-symmetry phase. Consequently, the energy landscape of ferroelectric Pca2$_1$ HfO$_2$ can be represented using these terms up to a sufficient order, and insight about the mechanism behind ferroelectricity can be extracted from it. The second-order terms are limited to simple quadratic or bilinear terms of the same irreps. In the third order, there are only trilinear terms that can be divided into two forms: $\Gamma \mathrm{X}_{\alpha} \mathrm{X}_{\alpha}$ couplings (involving one $\Gamma$-point mode and two X-point modes of the same wavevector) and $\mathrm{X}_{x} \mathrm{X}_{y} \mathrm{X}_{z}$ couplings (involving three X-point modes of different wavevectors). Similarly, quadlinear terms include $\Gamma \mathrm{X}_{x} \mathrm{X}_{y} \mathrm{X}_{z}$ and $\mathrm{X}_{\alpha} \mathrm{X}_{\alpha} \mathrm{X}_{\alpha'} \mathrm{X}_{\alpha'}$ categories. While there are other fourth order terms, they do not play an important role in reducing the symmetry, and hence we do not list them. All structural order parameters in Pca2$_1$ HfO$_2$ participate in odd-powered couplings with order parameters of other modes and must vanish in the reference structure \cite{zhao2025nature}.

\subsection{Minimal LGD Framework for Hybrid-Triggered Ferroelectricity}

With these insights, we now turn to investigate the microscopic origin of ferroelectricity in HfO$_2$. To demonstrate the mechanism by which trilinear and quadlinear couplings give rise to ferroelectricity, we first introduce a simplified, hypothetical model describing the electric enthalpy $H$ \cite{stengel2009electric, jung2024octahedral} in terms of a polar (along $\bm{c}$) and two nonpolar modes $p_0$ and $q_i$, all of which are stable. These quantities are distinct from the actual structural modes $P_0$ and $Q_i$ used to characterize HfO$_2$, but we consider the case where there is a trilinear coupling between them, similar to the case in HfO$_2$. This illustrative example captures the fundamental mechanism through which the trilinear coupling between between stable modes can generate a polar response. Building on this conceptual framework, we will extend the analysis to the real case of HfO$_2$, demonstrating that the same mechanism operates among its actual structural modes.
\begin{align}
    \begin{split}
        H &= E - \Omega \mathcal{E} P \\ 
          &= \frac{\beta_0}{2}{p_0}^2 + \frac{\beta_1}{2}{q_1}^2 + \frac{\beta_2}{2}{q_2}^2 + \gamma p_0q_1q_2 \\
          &+ \frac{\delta_0}{4}{p_0}^4 + \frac{\delta_{12}}{2}{q_1}^2{q_2}^2 - v(\lambda p_0)    
    \end{split} \label{simple}
\end{align}
The quantities $E$, $\Omega$, $\mathcal{E}$, and $P$ denote the energy, unit-cell volume, electric field, and polarization along the crystallographic direction $\bm{c}$, respectively. We expand the energy in terms of order parameters and adopt the reduced field variables  \cite{stengel2009electric}. $v$ represents the voltage across a unit cell along $\bm{c}$, and $\lambda$ corresponds to the mode effective charge \cite{gonze1997dynamical} associated with the polar mode, normalized by the lattice constant $c$. We consider the case where all coefficients are positive, thus there is no unstable mode that can give rise to ferroelectricity. The sign of $\gamma$ does not matter, since it is determined by the origin choice and the definition of the zone boundary order parameters.

Minimizing $H$ at fixed $v$ is equivalent to imposing a boundary condition of a fixed voltage applied along the $\bm{c}$ axis. The solutions of the following system of equations determine the equilibrium state of the material under the applied voltage, expressed in terms of the order parameters $p_0$, $q_1$, and $q_2$:
\begin{equation}
    \pdv{H}{p_0}=\pdv{H}{q_1}=\pdv{H}{q_2}=0
\end{equation}
We are interested in the existence of nontrivial solutions of this system, beyond the trivial case $q_1=q_2=0$. To this end, one may either solve $\partial H/\partial q_1=0$ for $q_1$ as a function of $p_0$ and $q_2$, or alternatively solve $\partial H/\partial q_2=0$ for $q_2$ as a function of $p_0$ and $q_1$. Adopting the latter approach yields
\begin{equation}
    q_2 = -\frac{\gamma p_0q_1}{\beta_2 + \delta_{12}{q_1}^2}
\end{equation}
Substituting this expression for $q_2$ into $\partial H/\partial q_1=0$, we obtain an effective condition involving only $p_0$ and $q_1$:
\begin{equation}
    \pdv{H}{q_1} = q_1 \left( \frac{\beta_1{\delta_{12}}^2{q_1}^4  + 2(\beta_1\beta_2\delta_{12}){q_1}^2 + \beta_1{\beta_2}^2 - \beta_2{\gamma }^2{p_0}^2}{{(\beta_2 + \delta_{12}{q_1}^2})^2} \right)
    \label{sup_dHdQ1_1_opt}
\end{equation}
The stability of the $q_1=0$ solution is determined by the sign of $\partial^2 H/\partial {q_1}^2$. If this quantity is negative at $q_1=0$, condensation of $q_1$ lowers the free energy, indicating the spontaneous condensation of the hybrid mode $q_1 q_2$. Conversely, a positive curvature implies that the mode remains inactive. Accordingly, the onset of the hybrid-mode instability is identified by the condition under which $\partial^2 H/\partial {q_1}^2$ evaluated at $q_1=0$ becomes negative:
\begin{equation}
       \left. \pdv[2]{H}{{q_1}} \right|_{q_1=0}
   = \frac{\beta_1{\beta_2}^2 - \beta_2{\gamma }^2{p_0}^2}{{\beta_2}^2} < 0
\end{equation}
This criterion defines the critical value of the order parameter $p_0$, denoted $p_{0,c}$, at which the instability occurs:
\begin{equation}
    |p_{0,c}| = \frac{\sqrt{\beta_1\beta_2}}{|\gamma |} \label{sup_P_0c}
\end{equation}

An illustrative solution of Equation~(\ref{simple}) is presented in \textbf{Figure~\ref{multicoupling}}a. Below  $|p_{\mathrm{0,c}}|$, the system behaves as a dielectric with no contribution to energy from the trilinear coupling. This region is shown with blue lines. However, once $|p_0|$ exceeds $|p_{\mathrm{0,c}}|$ by application of a voltage, the hybrid mode $q_1q_2$ becomes unstable inducing a  phase transition to the polar phase where both $q_1$ and $q_2$ condense simultaneously (shown in red lines)---similar to the ``avalanche'' transition in Aurivillius compounds \cite{etxebarria2010role}, and the predicted  transitions by quadratic-linear couplings \cite{gufan1980phenomenological}. The simplest explanation to this phenomenon is that $\gamma p_0$ acts as a tunable coefficient for second-order hybrid mode term $q_1q_2$. We refer to the pre-trigger phase as the dielectric phase and the post-trigger phase as the ``triggered'' phase.

\subsection{Triggered Ferroelectricity vs. Improper Ferroelectricity}

\begin{figure*}[t] 
	\centering
	\includegraphics[width=0.7\textwidth]{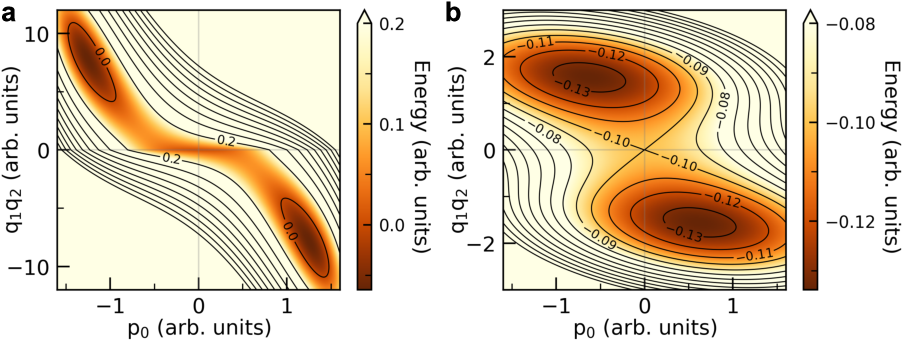} 

	\caption{\textbf{Coherent switching barrier of hybrid-triggered and hybrid improper ferroelectrics.}
    Example of an energy contour diagram from LGD model of 
    (a) Hybrid-triggered ferroelectrics from Equation~(\ref{simple}) and 
    (b) Hybrid improper ferroelectrics.
    The former switch though a barrier nonparallel to $p_0$ axis, while the latter encounter an intrinsic energy barrier parallel to the $p_0$ axis associated with the switching of either one of the nonpolar modes, which does not directly couple with voltage.
    }
	\label{contour}
\end{figure*}

A key distinction exists between improper and triggered ferroelectricity, though both arise from couplings between polar and nonpolar modes. Improper ferroelectricity involves unstable nonpolar modes (demonstrated by imaginary phonon frequency or negative quadratic coefficient of the order parameter)---either single or hybrid \cite{van2004origin, bousquet2008improper, Benedek2011}---coupled to the polar mode. 
As a result, polarization switching in improper ferroelectrics requires changing the direction of a nonpolar mode present in the structure, which introduces an energy barrier that does not directly couple with applied voltage. (\textbf{Figure~\ref{contour}}) Triggered ferroelectricity, by contrast, does not rely on unstable modes \cite{holakovsky1973new}. It arises from couplings of stable nonpolar modes with the polar mode (Figure S3). When polarization surpasses a critical threshold due to applied voltage, hybrid modes condense and the system moves to a different global minimum of energy. Unlike either proper and improper ferroelectricity, the mechanism underlying triggered ferroelectricity requires no additional instabilities.

This fundamental difference renders triggered ferroelectrics more useful for applications than improper ferroelectrics. For example, the energy barrier associated with switching of a nonpolar order parameter is absent in triggered ferroelectricity. The triggered mechanism also aligns with the observed dielectric increase upon heating, akin to a first-order transition between ferroelectric and dielectric phases \cite{gufan1980phenomenological}.
Notably, this is the only mechanism proposed so far that is compatible with the continuous evolution of antiferroelectricity to ferroelectricity in ZrO$_2$-based thin-films. $P$-$v$ plot of Figure~\ref{multicoupling}a shows existence of two distinct critical voltages, $v_d$ derived from the dielectric phase, and $v_t$ from the triggered phase. In the case where transition to the dielectric phase during polarization switching is merely transient ($v_d < v_t$), it results in a ferroelectric hysteresis. Conversely, if the dielectric phase persists and manifests as a non-transient state during polarization switching ($v_d > v_t$), it leads to the formation of an antiferroelectric-like double hysteresis.

\subsection{First-Principles Verification and Additional Couplings}

To connect the minimal LGD framework introduced above with the case of HfO$_2$,
we now explicitly map the abstract order parameters of Equation~(\ref{simple}) onto the
symmetry-adapted modes from section 2.1. In this mapping, the polar variable $p_0$
corresponds to the zone-center polar mode $P_0$, while pairs of nonpolar
variables $(q_1,q_2)$ represent symmetry-allowed combinations of
zone-boundary modes $(Q_i,Q_j)$ that participate in trilinear couplings of the
form $P_0 Q_i Q_j$. This correspondence enables a direct test
of the hybrid-triggered mechanism using first-principles calculations.

The modes $P_0$, $Q_1$, and $Q_5$ in HfO$_2$ exhibit symmetry properties consistent with those predicted by Equation~(\ref{simple}). (Analogous symmetry relationships are also found for $P_0Q_3Q_6$ and $P_0Q_3Q_7$; the $P_0Q_1Q_5$ trilinear coupling is selected here merely as a representative example. As will be shown, each of these coupling terms contributes to a single, unified ferroelectric mechanism.)
We set all other displacements to zero and compute the crystal's energy as a function of the hybrid mode $Q_1Q_5$, maintaining a fixed ratio between $Q_1$ and $Q_5$, while $P_0$ is held at progressively increasing values. The resulting energy plots are displayed in the left panel of Figure~\ref{multicoupling}c, where the critical value for triggered ferroelectricity is $|P_{\mathrm{0,c}}|$ is 0.49 Å. When $P_0$ is below this threshold, the hybrid mode increases the energy of HfO$_2$. For $P_0$ exceeding 0.49 Å, the energy of HfO$_2$ begins to decrease as a function of increasing hybrid mode order, revealing the hybrid mode instability.

Nevertheless, attributing the emergence of ferroelectricity to a single trilinear coupling term, $P_0 Q_1 Q_5$, proves insufficient, as the calculated critical value $|P_{\mathrm{0,c}}|$ exceeds the spontaneous polar distortion in the ferroelectric Pca2$_1$ phase, $|P_{\mathrm{0,s}}| = 0.40$~\AA. This discrepancy arises because the trilinear and quadlinear couplings that participate in the triggering mechanism of HfO$_2$ are interconnected in a manner that does not produce multiple, independent triggering events. Instead, these couplings act cooperatively to reduce the critical threshold of a single, unified trigger.

This behavior can be illustrated within the LGD framework. As an example, consider an extension of Equation~(\ref{simple}) in which an additional trilinear coupling term $q_1 q_3 q_4$ is introduced, together with higher-order terms to ensure stability:
\begin{align}
    \begin{split} \label{eq:multicoupling}
        H &= \frac{\beta_0}{2}{p_0}^2 + \frac{\beta_1}{2}{q_1}^2 + \frac{\beta_2}{2}{q_2}^2 + \frac{\beta_3}{2}{q_3}^2 + \frac{\beta_4}{2}{q_4}^2 \\
        &+ \gamma_{012}p_0q_1q_2 + \gamma_{134}q_1q_3q_4 + \frac{\delta_0}{4}{p_0}^4 + \frac{\delta_4}{4}{q_4}^4 \\
        &+ \frac{\delta_{12}}{2}{q_1}^2{q_2}^2 + \frac{\delta_{13}}{2}{q_1}^2{q_3}^2 - v(\lambda p_0)
    \end{split}
\end{align}
Here, $q_4$ corresponds to the order parameter of an unstable mode, implying the condition $\beta_4 < 0$. The couplings $p_0 q_1 q_2$ and $q_1 q_3 q_4$ each involve one mode that condenses in the dielectric phase ($p_0$, which couples directly to the applied voltage, and $q_4$, which is intrinsically unstable) and two modes that condense only in the triggered phase ($q_1 q_2$ and $q_1 q_3$). Crucially, both couplings share a common component that remains inactive prior to triggering, namely $q_1$. A representative solution of Equation~(\ref{eq:multicoupling}) is shown in Figure~\ref{multicoupling}b. Compared to Equation~(\ref{simple}), this extended model exhibits a unified triggering event in which $p_0$, $q_1$, and $q_2$ condense simultaneously at a lower critical polarization than that predicted by Equation~(\ref{sup_P_0c}). The detailed derivation is provided in the Supporting Information.

An analogous situation occurs in HfO$_2$, where all relevant couplings collectively drive a single transition in which every mode that is inactive in the dielectric phase condenses at a unified and reduced critical value $|P_{\mathrm{0,c}}|$. Figure~\ref{modes_coupling}b illustrates how nearly all trilinear and quadlinear couplings contribute to the triggering mechanism in HfO$_2$, with the exceptions of $P_0 Q_2 Q_4$, $Q_1 Q_3 Q_5 Q_6$, and $Q_1 Q_3 Q_5 Q_7$. In the right panel of Figure~\ref{multicoupling}c, we present DFT energies computed for the same fixed values of $P_0$, $Q_1$, and $Q_5$ as in the left panel, but now allowing all other order parameters to relax to nonzero values. This relaxation leads to a substantial reduction of the critical polarization, lowering $P_{\mathrm{0,c}}$ from 0.49~\AA\ to 0.17~\AA.

\begin{figure*}[t] 
	\centering
	\includegraphics[width=0.8\textwidth]{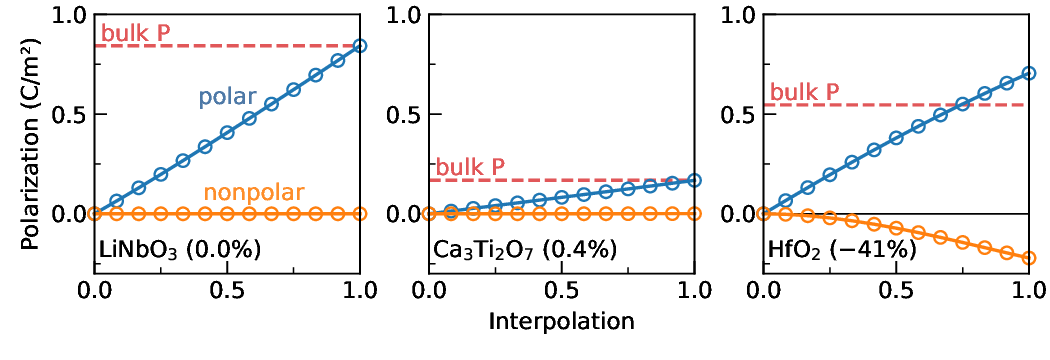} 

    \caption{\textbf{Higher-order dynamical charge arising from nonpolar modes in HfO$_2$.} 
    Interpolation between the high-symmetry reference structure and the polar ground state configuration, decomposed into polar and nonpolar mode contributions. 
    HfO$_2$ exhibits a substantial higher-order polarization component originating from nonpolar modes (accounting for approximately --41\% of the bulk response), 
    in sharp contrast to the proper ferroelectric LiNbO$_3$, where such contributions are symmetrically forbidden (0.0\%), and to hybrid improper ferroelectric Ca$_3$Ti$_2$O$_7$, 
    where they are allowed but negligible (0.4\%).
    }
	\label{interpol}
\end{figure*}

\subsection{Higher-Order Dynamical Charges}

Even after addressing the roles of multiple trilinear and quadlinear couplings, an important aspect of this transition remains unexplained. As demonstrated in previous studies \cite{zhou2022strain,qi2025sign}, when the DFT energy of HfO$_2$ is accurately plotted as a function of the polar order parameter, a cusp emerges in the curve. This cusp is does not appear in our demonstrations of hybrid-triggered ferroelectricity so far, as it is not a feature necessarily derived from the mechanism itself. A major contribution to this energy cusp is from an underappreciated factor which also connects to the origin of strong trilinear couplings in HfO$_2$: the polarization arising from hybrid modes which are nonpolar by themselves. The hybrid modes $Q_1Q_5$, $Q_2Q_4$, $Q_3Q_6$, and $Q_3Q_7$, which couple with the polar mode in HfO$_2$, break all the inversion symmetries present in the crystal and transforms exactly as the polarization itself ($\Gamma_{4}^-$) (Table~S2). 

The broken inversion symmetry from nonpolar modes is a well-established characteristic of hybrid improper ferroelectrics \cite{Benedek2011, harris2011symmetry}. In such systems, the coupling between two nonpolar modes typically induces a nonzero amplitude of the polar structural order parameter, which serves as the dominant source of polarization. However, HfO$_2$ deviates from this conventional picture. In \textbf{Figure~\ref{interpol}}, the ground state structures of the proper ferroelectric LiNbO$_3$ (R3c, \#161), the hybrid improper ferroelectric Ca$_3$Ti$_2$O$_7$ (Cmc2$_1$, \#36), and HfO$_2$ (Pca2$_1$, \#29) are each decomposed into two components: one arising from distortions from polar modes and the other from nonpolar modes. Each component is interpolated with its respective high-symmetry parent phase (Pm$\bar{3}$m, I4/mmm, and Fm$\bar{3}$m), and the resulting polarization is plotted. As expected, the nonpolar distortion of LiNbO$_3$ produces no polarization, while the hybrid improper Ca$_3$Ti$_2$O$_7$ exhibits only a minute polarization induced through the trilinear coupling that enables its ferroelectricity.

In contrast, the interpolated structures of HfO$_2$ exhibit a distinctly different behavior, highlighting the critical influence of both trilinear and quadlinear couplings on the emergence of its ferroelectricity. The nonpolar distortions produce a nonlinear polarization response, contributing up to approximately --41\% of the total bulk polarization. These results indicate that, unlike in conventional improper or hybrid improper ferroelectrics, the direct coupling of hybrid modes—or, equivalently, higher-order dynamical charge interactions with the electric field---together with local electrostatic effects, plays a decisive role in the ferroelectric behavior of HfO$_2$. This interpretation is further supported by recent reports of ion motion that reverses the direction of the simulated electric field \cite{choe2021unexpectedly,qi2025polarization,yang2025theoretical}. a phenomenon that cannot be simply explained within the traditional framework of linear dynamical charges (Born effective charges).

\begin{figure*}[t] 
	\centering
	\includegraphics[width=0.90\textwidth]{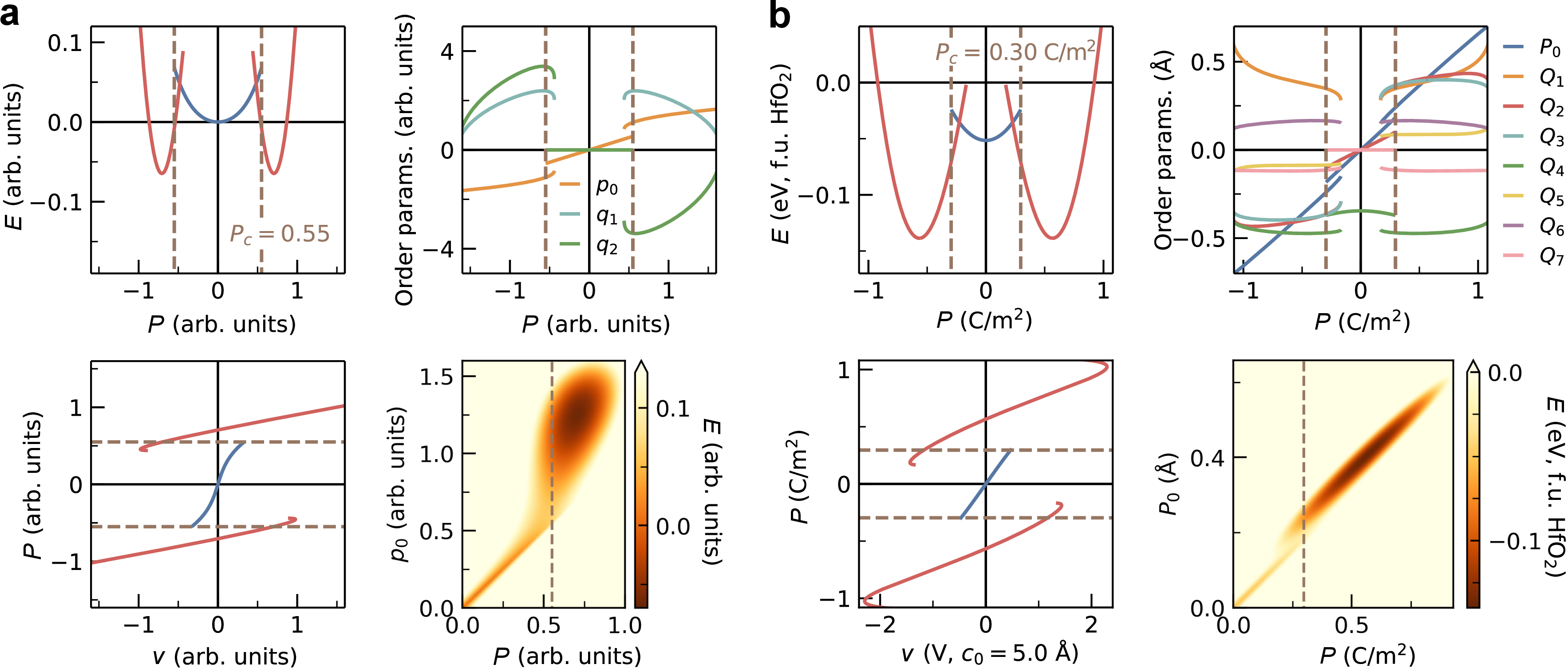} 

	\caption{\textbf{The effect of polar hybrid modes on triggered ferroelectricity.}
        (a) Plots for a hybrid-triggered ferroelectric material with polar hybrid mode, including energy-polarization (top left), polarization-voltage (bottom left), order parameters-polarization (top right), and two-dimensional energy-$p_0$,$P$ (bottom right) modeled with additional hybrid mode contribution to polarization to Equation~(\ref{simple}). The triggered ferroelectric phase is separated from the dielectric phase in the two-dimensional energy plot, introducing an additional layer of hysteresis and a disruptive phase transition.
        (b) Coherent switching pathway of HfO$_2$ via the hybrid-triggered mechanism. Five order parameters that remain zero in the dielectric phase condense together at $P_\mathrm{c} = 0.30$ C\,m$^{-2}$ disruptively.
        }
	\label{charge_effect} 
\end{figure*}

Accordingly, a new term representing the electronic polarization induced by the hybrid mode must be added to Equation~(\ref{simple}), such that $\lambda p_0 \rightarrow \lambda p_0 +\mu q_1q_2$.
\begin{align}
\begin{split}
     H& = \frac{\beta_0}{2}{p_0}^2 + \frac{\beta_1}{2}{q_1}^2 + \frac{\beta_2}{2}{q_2}^2 + \gamma p_0q_1q_2 \\
     &+ \frac{\delta_0}{4}{p_0}^4 + \frac{\delta_{12}}{2}{q_1}^2{q_2}^2 - v(\lambda p_0+\mu q_1q_2)
\end{split}
\end{align}
Here, $\mu$ acts as normalized second-order mode effective charge, derived from the second derivative of polarization with respect to nonpolar atomic displacements \cite{jung2025electric}. Such a variable is not commonly introduced before. Solution of the resulting equation with the same coefficients as Equation~(\ref{simple}) is shown in \textbf{Figure~\ref{charge_effect}}a. The polarization of the hybrid mode is against the direction of the polar mode increases the value of $P_{\mathrm{0,c}}$. This addition generates the cusp in the energy vs polarization plot and reveals more aspects of the phase transition. Not only are there overlapping regions of polarization in the transition pathway, but also the voltage and the order parameters in these overlapping regions are discontinuous, resulting in a disruptive phase transition where all the order parameters abruptly shift as polarization increases.

The energy plotted as a function of $P_0$ and $P$  provides further insight into this transition. (Figure~\ref{charge_effect}a, bottom right). When $\mu$ is zero, the data in this plot is restricted to a one-dimensional line, as $P$ cannot assume a value different from $\lambda P_0$, and vice versa. However, when $\mu$ is non-zero, $P$ and $P_0$ are independent. Two distinct local minima emerge in this plot: the dielectric phase where the hybrid mode is inactive, and the triggered phase where they condense. As $\mu$ increases, the triggered phase's local minimum separates more from the $P = \lambda P_0$ line, eventually disconnecting entirely from the dielectric phase's local minimum. Thus, a slight increase in polarization above $P_c$ causes a sudden jump in all order parameters, including $P_0$. The pathway from the triggered phase to the dielectric phase differ from the pathway from the dielectric phase to the triggered phase as the local minima are completely separated, leading to overlapping polarization regions and additional hysteresis in the switching pathway.

\subsection{Full Energy Landscape and Coherent Switching Pathways}

We apply these insights to HfO$_2$ under 1\% epitaxial tensile strain to map the complete coherent ferroelectric phase transition pathway. Note that switching in the real material does not necessarily happen in a coherent, homogeneous way, and what we discuss here is not an exact reproduction of factors such as domain wall motion which likely affect the switching energy barriers \cite{ye2025ab}. However, what we present is a complete energy landscape of most commonly observed Pca2$_1$ HfO$_2$ that involves all possible parent structure of Pca2$_1$ which is also a subgroup of I4/mmm (strained Fm$\bar{3}$m). This understanding is crucial for interpreting the complex nature of real switching behavior and domain boundaries, since even materials with intricate domain structures exhibit coherent switching in the local limit.

We derive the symmetry-adapted expansion of the energy up to fourth order and of the polarization up to second order (Equation~(S3)) for HfO$_2$, and fit these expressions to DFT-computed energy and polarization data sampled over eight-dimensional grid in order parameter space around the I4/mmm and Pca2$_1$ phases of HfO$_2$ (Figure S4). The resulting solutions obtained from the fitted coefficients are presented in Figure~\ref{charge_effect}b. The critical polarization is 0.30 C\,m$^{-2}$. In the dielectric phase without applied voltage, only $Q_4$ is non-zero; upon applying voltage, $P_0$ and $Q_2$ condense. At 0.47 V/unit cell (u.c.), ferroelectricity is triggered, causing a simultaneous jump in five other order parameters. The voltage needed to switch polarization is higher than the triggering voltage, reaching about 1.4 V/u.c.. The separation of two local minima is clearly visible in the two-dimensional energy landscape. Note that the proper ferroelectric switching pathway via the Pcam (\#57) phase is also observed from this model, but requires higher voltage of 2.3 V/u.c., and is geometrically restricted in non-periodic boundary conditions. Thus, it is reasonable to conclude that HfO$_2$ will switch through the Aeaa phase via the hybrid-triggered mechanism under homogeneous conditions.

\begin{figure*}[t] 
	\centering
	\includegraphics[width=0.67\textwidth]{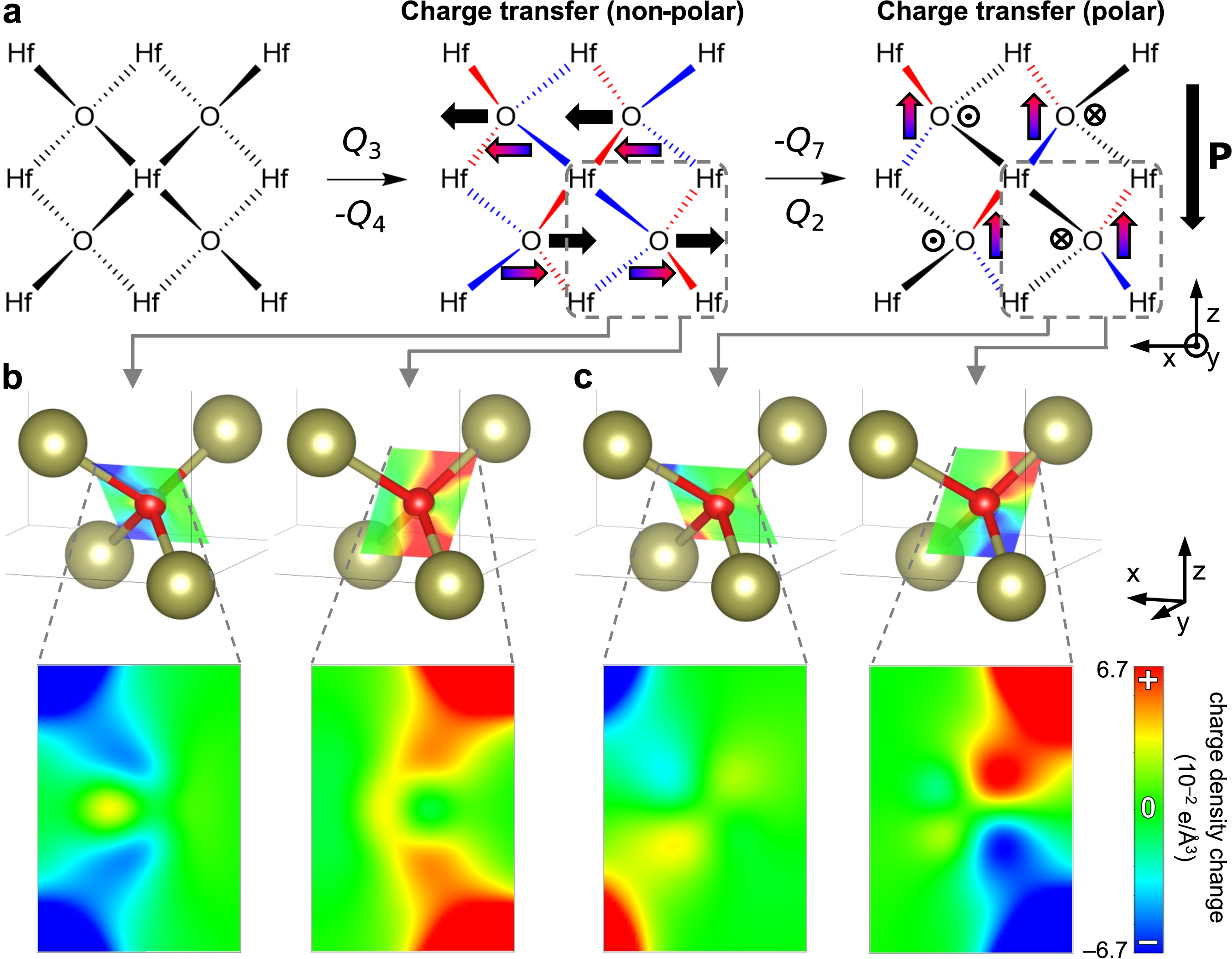}

	\caption{\textbf{The origin of polar hybrid modes in fluorite structure.}
        (a) A (010) plane of oxygens ($y=0.75$) and surrounding Hf atoms in fluorite HfO$_2$ unit cell with gradient-colored arrows indicating charge transfer from the elongated Hf-O bonds (blue) to the shortened bonds (red). 
        (b) Charge transfer resulting from the $-Q_4$ or $Q_3$ mode around highlighted oxygen ion. The plane containing the two bonds directed towards $+x$ shows charge loss (left), whereas the plane containing the two bonds directed towards $-x$ shows charge gain.
        (c) Charge transfer resulting from the $Q_2$ or $-Q_7$ mode starting from condensed $-Q_4$ or $Q_3$ mode. There is minimal charge transfer between the two extended bonds in the $+x$ direction (left), while there is significant charge transfer between the two shortened bonds in the $-x$ direction resulting in net polar structure.
    }
	\label{hybrid_charge} 
\end{figure*}

\subsection{Origin of Polar Hybrid Modes from Nonpolar Modes}

Finally, one critical question remains: what makes the hybrid modes so polar? The answer lies more in the fluorite structure itself rather than in a unique property of the Hf-O bonds. In fact, the Hf-O bond is relatively simple electronically due to the high electropositivity of Hf, compared to bonds such as Ti-O which are known to enable ferroelectricity in perovskites through hybridization \cite{cohen1992origin}. Although the Hf-O bond is mostly ionic, it gains more covalent character as the Hf-O bond length decreases. This results in bond-to-bond charge transfer associated with ionic displacements in each mode \cite{jung2025electric}. \textbf{Figure~\ref{hybrid_charge}}a shows a (010) plane of oxygens ($y=0.75$) and surrounding Hf atoms in the conventional cell. When ions displace according to $-Q_4$ or $Q_3$, charge transfers towards $+x$ between the Hf-O bonds in the upper half, and towards $-x$ in the lower half. Figure~\ref{hybrid_charge}b shows this charge transfer from DFT, in the form of cuts of charge density change on Hf-O planes. The opposing direction of charge transfer between bonds results in a overall nonpolar structure. 

However, if ions then displace according to $Q_2$ or $-Q_7$ starting from the distorted structure by $-Q_4$ or $Q_3$, charge is transferred upward around each oxygen ion in the crystal, as the charge transfer from the shortened bonds surpasses that from the elongated bonds (Figure~\ref{hybrid_charge}c). This results in a significant downward polarization purely from electronic dipole effects, without any ionic contribution \cite{jung2025electric}. This aligns with the observation that the Born effective charges are dramatically different in different phases \cite{fan2022vibrational}. The nontrivial polarization in this system originate less from the first derivatives of polarization with respect to displacements, but more from the second derivatives. Also this purely electronic dipole, coupled with an oppositely oriented ionic dipole, can enhance the stability of ferroelectricity in ultra-thin HfO$_2$, as the depolarization field from one polarization source stabilizes the other, further strengthening the trilinear interaction.

\section{Conclusions}

Our findings transform the traditional interpretation of polarization in oxides as a linear function of charged ions' displacements, especially in HfO$_2$. HfO$_2$ was thought to consist of nonpolar spacer layers and polar layers, based on the displacement of oxygen ions. We instead show that HfO$_2$ can exhibit substantial polarization $P_z$ even in the absence of ionic displacements along $z$. We term this newly identified mechanism of ferroelectricity, which arises from strong multilinear couplings among stable modes, \textit{hybrid-triggered ferroelectricity}. This phenomenon is conceptually analogous to triggered ferroelectricity \cite{holakovsky1973new}, in the same way that hybrid improper ferroelectricity \cite{Benedek2011} relates to improper ferroelectricity \cite{levanyuk1974improper}. These insights resolve longstanding debate regarding the ferroelectric behavior of thin-film HfO$_2$, and also establish a conceptual framework for designing next-generation ferroelectric materials and devices.

\section{Computational Methods}
Periodic density functional theory (DFT) calculations were performed with the Vienna ab initio simulation package (VASP) 6.4.1 \cite{kresse1996efficient}. Spin-orbit coupling as well as spin-polarization was tested to contribute no qualitative difference to the system. A generalized gradient approximation (GGA) exchange-correlation functional PEBsol was used \cite{perdew2008restoring}. The projector augmented wave (PAW) method was used to describe atom cores and the plane wave basis set was expanded to a kinetic energy maximum of 520 eV for Kohn-Sham orbitals, using valence configurations 5$p^6$5$d^3$6$s^1$ for Hf and 2$s^2$2$p^4$ for O. Tetrahedron smearing with Bl\"{o}chl corrections imposed on electrons at the Fermi level. VESTA software \cite{momma2011vesta} was used for building and visualizing crystal structures. More details on the computational method can be found in Supporting Information.

\printbibliography
\end{refsection} 

\section*{Acknowledgments}
\paragraph*{Funding:}
This work is supported by the Office of Naval Research Grant N00014-24-1-2082. The authors acknowledge the Minnesota Supercomputing Institute (MSI) at the University of Minnesota for providing part of the computational resources that contributed to this work. 

\paragraph*{Author contributions:}
S.J. conceived the idea, designed, performed the calculations and analyzed data. T.B. supervised the research. S.J. prepared the manuscript. S.J. and T.B. edited the manuscript.
\paragraph*{Conflicts of interests:}
There are no conflicts of interests to declare.
\paragraph*{Data and materials availability:}
Data not presented in the manuscript is available from the authors upon request.

\end{multicols}


\clearpage
\newpage
\renewcommand{\thefigure}{S\arabic{figure}}
\renewcommand{\thetable}{S\arabic{table}}
\renewcommand{\theequation}{S\arabic{equation}}
\renewcommand{\thepage}{S\arabic{page}}
\setcounter{figure}{0}
\setcounter{table}{0}
\setcounter{equation}{0}
\setcounter{page}{1} 

\begin{center}
\section*{Supporting Information for\\ \scititle}

Seongjoo~Jung,
Turan~Birol$^\ast$\\ 
\small$^\ast$Corresponding author. Email: tbirol@umn.edu\\
\end{center}

\begin{refsection}
\subsection*{Computational Method}

7 $\times$ 7 $\times$ 7 $\Gamma$-centered \textbf{k}-point mesh was used to sample the Brillouin zone of the 12-atom supercell of fluorite HfO$_2$. Optimization of structures was converged below 10$^{-7}$ eV of electronic energy changes and 1.0 meV/\AA~of maximum ionic forces. Fixed-ion calculations were converged below 10$^{-6}$ eV of energy changes. Lattice parameter for cubic HfO$_2$ was 5.020 Å. Biaxial strain was simulated by fixing lattice vectors $\bm{a}$, $\bm{b}$ and allowing $\bm{c}$ to relax. (The $\eta_{zz}$ strain has same irrep as energy for a parent structure I4/mmm, appearing implicit in the Landau equation \cite{jung2024octahedral}). Lattice parameters under 1\% tensile strain were $a=b=5.070$ Å, $c_0=5.002$ Å for I4mmm phase, $c=5.038$ Å for Pca2$_1$ phase, $c=5.112$ Å for P4$_2$/nmc phase and $c=5.020$ Å for Aeaa phase.
Polarization was calculated using the Berry phase definition from modern theory of polarization \cite{king1993theory}. 15 $\times$ 15 $\times$ 15 $\Gamma$-centered \textbf{k}-point mesh, and 160 $\times$ 160 $\times$ 160 fine real space mesh was used for charge density representation of the 12-atom cubic supercell.

A 30-atom and a 48-atom supercell were used to simulate the ground state structures of LiNbO$_3$ and Ca$_3$Ti$_2$O$_7$, corresponding to the space groups R$\bar{3}$c (\#161) and Cmc2$_1$ (\#36), respectively. For LiNbO$_3$, a $\Gamma$-centered $8 \times 8 \times 3$ \textbf{k}-point mesh and a plane-wave kinetic energy cutoff of 650 eV were employed, while for Ca$_3$Ti$_2$O$_7$, a $\Gamma$-centered $6 \times 6 \times 2$ \textbf{k}-point mesh and a cutoff energy of 520 eV were used. The valence electron configurations adopted were $1s^22s^1$ for Li, $4s^24p^64d^45s^1$ for Nb, $3s^23p^64s^2$ for Ca, and $3s^23p^63d^34s^1$ for Ti. The optimized lattice parameters of the ground state structures were $a = b = 5.13$ Å, $c = 13.80$~Å, $\gamma = 120^{\circ}$ for LiNbO$_3$; $a = 5.44$ Å, $b = 5.39$ Å, $c = 19.31$ Å for Ca$_3$Ti$_2$O$_7$; and $a = 5.20$ Å, $b = 5.00$ Å, $c = 5.02$ Å for the HfO$_2$ supercells. The space groups of each ground structures separated into polar and non-polar modes were R3m (\#160) and R$\bar{3}$c (\#167) for LiNbO$_3$, Fmm2 (\#42) and Cmc2$_1$ (\#36) for Ca$_3$Ti$_2$O$_7$, and Fmm2 (\#42) and Pca2$_1$ (\#21) for HfO$_2$.

Space group representations, mode definitions and couplings were referred from the Bilbao
Crystallographic Server \autocite{aroyo2011crystallography} and FINDSYM, ISODISTORT, and INVARIANTS from the ISOTROPY software suite \autocite{hatch2003invariants,stokes2005findsym,campbell2006isodisplace,stokes_isotropy}. All modes present in both the polar and nonpolar minima was be considered. (The condensed modes in the triggered phase also include those in the dielectric phase.) Additionally, we assume that no additional mode condenses exclusively between the two minima, which is a valid assumption for macroscopic modeling of ferroelectricity without domain separation where no net polarization should appear in the in-plane direction without electrode screening. The following order parameter definition involving normalization factor $\sqrt{N}$ was used, where $N$ is the number of primitive cells in the u.c. and $u$ is the displacement of an ion based on the lattice parameters of high-symmetry I4/mmm structure  \autocite{cowley1980structural}:
\begin{gather}
    |Q|=\sqrt{\frac{\Sigma {u_{i\alpha}}^2}{N}}
\end{gather}
$i$ denotes iteration over each ion in the u.c., and $\alpha$ over each Cartesian directions. 

\begin{figure}
    \centering
    \includegraphics[width=0.6\textwidth]{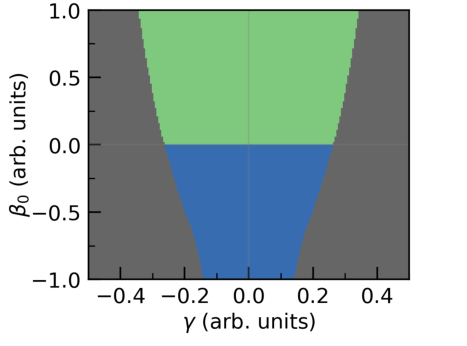} 
    \caption{\textbf{Phase diagram of a hybrid-triggered ferroelectric by eq.~\ref{simple}, by second-order $p_0$ coefficient $\beta_0$ and trilinear coupling coefficient $\gamma$.} 
    Green area represent the nonpolar ground state where $p_0=q_1=q_2=0$, blue area the proper ferroelectric ground state where $p_0 \neq 0$, $q_1=q_2=0$, and the gray area the hybrid-triggered polar ground state where $p_0,q_1,q_2 \neq 0$. The hybrid-triggered ground state exists for cases when the polar mode is both soft and hard, if the trilinear coupling is strong enough. Similar phase diagram can be found for a hypothetical case of quadratic-linear order parameter coupling \autocite{gufan1980phenomenological}, and charge density wave order in kagome metals involving a trilinear coupling \autocite{christensen2021theory}. The tricritical points are along $\beta_0=0$ for all phase diagrams. In the cases where the components of hybrid mode remain hard when $\beta_0=0$, there is a tricritical point for each sign of non-zero value of $\gamma$. In the case of kagome metals where all three components of trilinear coupling have the same second-order coefficient, the two tricritical points merge into one point at $\gamma=0$.} \label{sup_phase}
\end{figure}

The equation and coefficients used for the figures are as follows, which are unitless and chosen for illustrative purposes: \\
Fig.~\ref{multicoupling}a, \ref{contour}a: Equation (1), \\ $(\beta_0,\beta_1,\beta_2,\gamma,\delta_0,\delta_{12},\lambda)=(0.3,0.2,0.1,0.3,1,0.03,1)$. \\
Fig.~\ref{multicoupling}b: Equation (7), \\ $(\beta_0,\beta_1,\beta_2,\beta_3, \beta_4,\gamma_{012},  \gamma_{134},\delta_0,\delta_{12}, \delta_{13}, \delta_4,\lambda) = (0.3,0.2,0.1, 0.1, -0.4,0.3, 0.25,1,0.03, 0.12, 1.6,1)$. \\
Fig.~\ref{contour}b: 
\begin{align}
    H = \frac{\beta_0}{2}{p_0}^2 + \frac{\beta_1}{2}{q_1}^2 + \frac{\beta_2}{2}{q_2}^2 + \gamma p_0q_1q_2 + \frac{\delta_1}{4}{q_1}^4 + \frac{\delta_2}{4}{q_2}^4 + - v(\lambda p_0)
\end{align}
$(\beta_0,\beta_1,\beta_2,\gamma,\delta_1,\delta_{2},\lambda)=(0.05,-0.2,-0.1,0.02,0.1,0.1,1)$. \\
Fig.~\ref{charge_effect}a: Equation (8), \\ $(\beta_0,\beta_1,\beta_2,\gamma,\delta_0,\delta_{12},\lambda,\mu)=(0.3,0.2,0.1,0.3,1,0.03,1,0.07)$. \\
Fig.~\ref{sup_phase}: Equation (1), $(\beta_1,\beta_2,\delta_0,\delta_{12},\lambda,v)=(0.2,0.1,1,0.03,1,0)$.

\begin{figure}
    \centering
    \includegraphics[width=0.65\textwidth]{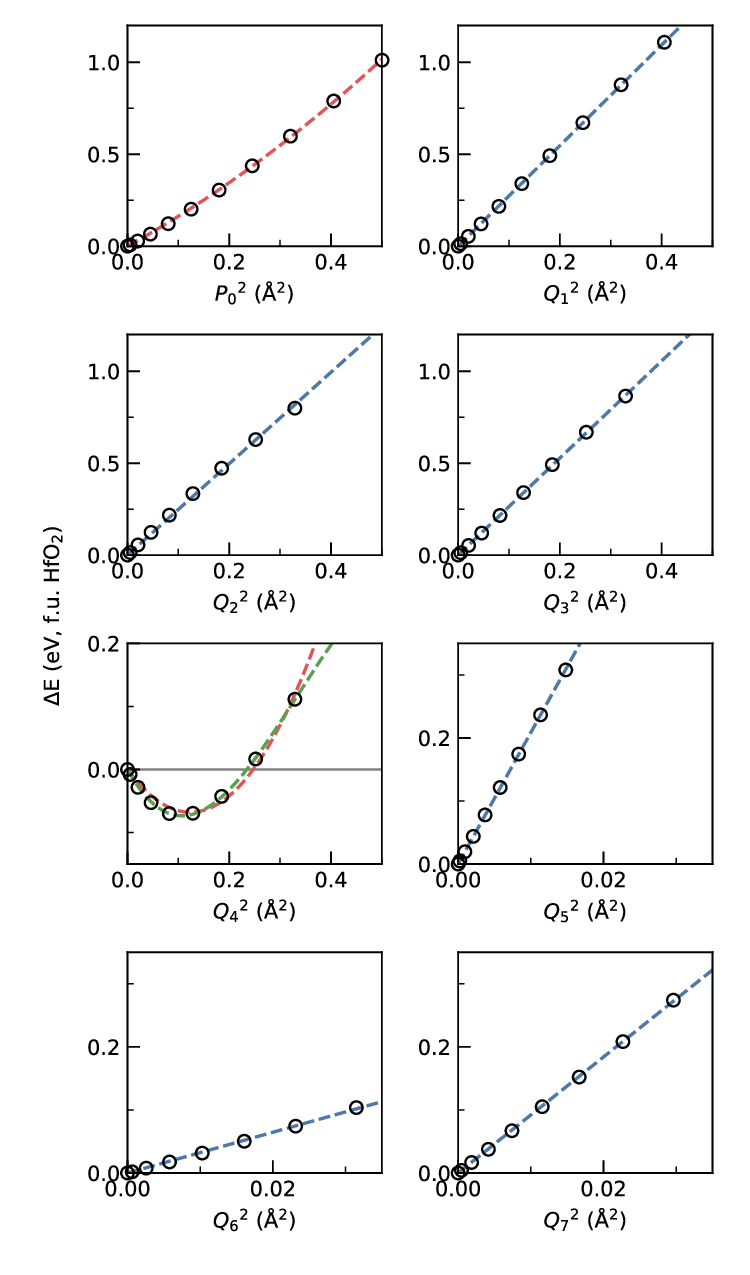} 
    \caption{\textbf{Contribution of each mode to energy from DFT.} 
    Blue, red and green dotted lines each represent 1st, 2nd and 3rd order regression of energy with respect to $Q^2$ through the origin. Linear regression fits most order parameters squared, except $P_0$ and $Q_4$.
    } \label{sup_1d}
\end{figure}

The full equation and coefficients used to represent entire order parameter space of HfO$_2$ in Fig.~\ref{charge_effect}b,c is:
\begin{align} \label{full}
    \begin{split}
        H &= \beta_0{P_0}^2 + \beta_1{Q_1}^2 + \beta_2{Q_2}^2 + \beta_3{Q_3}^2 + \beta_4{Q_4}^2 + \beta_5{Q_5}^2 + \beta_6{Q_6}^2 + \beta_{67}{Q_6}{Q_7} + \beta_7{Q_7}^2 \\
       &+ \gamma_{015}{P_0}{Q_1}{Q_5} + \gamma_{024}{P_0}{Q_2}{Q_4} + \gamma_{036}{P_0}{Q_3}{Q_6} + \gamma_{037}{P_0}{Q_3}{Q_7}  \\
       &+ \gamma_{123}{Q_1}{Q_2}{Q_3} + \gamma_{146}{Q_1}{Q_4}{Q_6} + \gamma_{147}{Q_1}{Q_4}{Q_7} + \gamma_{256}{Q_2}{Q_5}{Q_6} + \gamma_{257}{Q_2}{Q_5}{Q_7} + \gamma_{345}{Q_3}{Q_4}{Q_5} \\
       &+ \delta_{0}{P_0}^4 + \delta_{01}{P_0}^2{Q_1}^2 + \delta_{02}{P_0}^2{Q_2}^2 + \delta_{12}{Q_1}^2{Q_2}^2 + \delta_{03}{P_0}^2{Q_3}^2 + \delta_{13}{Q_1}^2{Q_3}^2 + \delta_{23}{Q_2}^2{Q_3}^2 + \delta_{04}{P_0}^2{Q_4}^2 + \delta_{14}{Q_1}^2{Q_4}^2  \\
       &+ \delta_{24}{Q_2}^2{Q_4}^2 + \delta_{34}{Q_3}^2{Q_4}^2 + \delta_{4}{Q_4}^4 + \delta_{05}{P_0}^2{Q_5}^2 + \delta_{15}{Q_1}^2{Q_5}^2 + \delta_{25}{Q_2}^2{Q_5}^2 + \delta_{35}{Q_3}^2{Q_5}^2 + \delta_{45}{Q_4}^2{Q_5}^2 \\
       &+ \delta_{06}{P_0}^2{Q_6}^2 + \delta_{16}{Q_1}^2{Q_6}^2 + \delta_{26}{Q_2}^2{Q_6}^2 + \delta_{36}{Q_3}^2{Q_6}^2 + \delta_{46}{Q_4}^2{Q_6}^2 + \delta_{56}{Q_5}^2{Q_6}^2 \\
       &+ \delta_{067}{P_0}^2{Q_6}{Q_7} + \delta_{167}{Q_1}^2{Q_6}{Q_7} + \delta_{267}{Q_2}^2{Q_6}{Q_7} + \delta_{367}{Q_3}^2{Q_6}{Q_7} + \delta_{467}{Q_4}^2{Q_6}{Q_7} + \delta_{567}{Q_5}^2{Q_6}{Q_7} + \delta_{667}{Q_6}^3{Q_7} \\
       & + \delta_{67}{Q_6}^2{Q_7}^2 + \delta_{07}{P_0}^2{Q_7}^2 + \delta_{17}{Q_1}^2{Q_7}^2 + \delta_{27}{Q_2}^2{Q_7}^2 + \delta_{37}{Q_3}^2{Q_7}^2 + \delta_{47}{Q_4}^2{Q_7}^2 + \delta_{57}{Q_5}^2{Q_7}^2 + \delta_{677}{Q_6}{Q_7}^3\\
       &+ \delta_{0126}{P_0}{Q_1}{Q_2}{Q_6} + \delta_{0127}{P_0}{Q_1}{Q_2}{Q_7} + \delta_{0134}{P_0}{Q_1}{Q_3}{Q_4} + \delta_{0235}{P_0}{Q_2}{Q_3}{Q_5} + \delta_{0456}{P_0}{Q_4}{Q_5}{Q_6} + \delta_{0457}{P_0}{Q_4}{Q_5}{Q_7} \\
       &+ \delta_{1245}{Q_1}{Q_2}{Q_4}{Q_5} + \delta_{1356}{Q_1}{Q_3}{Q_5}{Q_6} + \delta_{1357}{Q_1}{Q_3}{Q_5}{Q_7} + \delta_{2346}{Q_2}{Q_3}{Q_4}{Q_6} + \delta_{2347}{Q_2}{Q_3}{Q_4}{Q_7} \\
       &-v(\lambda P_0 + \mu_{15}Q_1Q_5 + \mu_{24}Q_2Q_4 + \mu_{36}Q_3Q_6 + \mu_{37}Q_3Q_7)
    \end{split}
\end{align}
Unlike others such as eq.~\ref{simple}, this equation does not include conventional factors of $1/n$ for the coefficients of $Q^n$ terms. Fourth order terms ${Q_1}^4,{Q_2}^4,{Q_3}^4,{Q_5}^4,{Q_6}^4,{Q_7}^4$ which have minimal effect in the energy (Fig.~\ref{sup_1d}) has been excluded from the full regression up to 4th order for better convergence in the $P$-$P_0$ plot.

DFT data used for regression involves combinations of following order parameters. Around I4/mmm: \\
$P_0=(-0.0707,0,0.0707)$ Å, 
$Q_1=(-0.0707,0,0.0707)$ Å, 
$Q_2=(-0.0717,0,0.0717)$ Å, 
$Q_3=(-0.0717,0,0.0717)$ Å, \\
$Q_4=(-0.0717,0,0.0717)$ Å, 
$Q_5=(-0.0152,0,0.0152)$ Å, 
$Q_6=(-0.0254,0,0.0254)$ Å, 
$Q_7=(-0.0215,0,0.0215)$ Å. \\
Around Pca2$_1$:
$P_0=(0.3299, 0.4007, 0.4714)$ Å,
$Q_1=(0.3088, 0.3795, 0.4503)$ Å,
$Q_2=(0.3283, 0.4001, 0.4718)$ Å,\\
$Q_3=(0.3103, 0.3820, 0.4537)$ Å,
$Q_4=(-0.5411, -0.4694, -0.3977)$ Å,
$Q_5=(0.0708, 0.0860, 0.1013)$ Å, \\
$Q_6=(0.1344, 0.1597, 0.1851)$ Å,
$Q_7=(-0.1398, -0.1182, -0.0967)$ Å.


\subsection*{Derivation of hybrid-triggered instability with additional couplings}

\begin{align}
    \begin{split} \label{sup_H1}
        H &= \frac{\beta_0}{2}{p_0}^2 + \frac{\beta_1}{2}{q_1}^2 + \frac{\beta_2}{2}{q_2}^2 + \frac{\beta_3}{2}{q_3}^2 + \frac{\beta_4}{2}{q_4}^2 + \gamma_{012}p_0q_1q_2 + \gamma_{134}q_1q_3q_4  \\
        &+ \frac{\delta_0}{4}{p_0}^4 + \frac{\delta_4}{4}{q_4}^4 + \frac{\delta_{12}}{2}{q_1}^2{q_2}^2 + \frac{\delta_{13}}{2}{q_1}^2{q_3}^2 - v(\lambda p_0)
    \end{split}
\end{align}
Note that $\beta_4 < 0$, as $q_4$ corresponds to the order parameter of an unstable mode. The derivation follows the three steps presented in the main text. First, determine the partial derivatives.
\begin{align}
    \pdv{H}{p_0} &= \beta_0{p_0} + \gamma_{012}q_1q_2 + \delta_0{p_0}^3 - \lambda v = 0 \\
    \pdv{H}{q_1} &= \beta_1{q_1} + \gamma_{012}p_0q_2 + \gamma_{134}q_3q_4 + \delta_{12}q_1{q_2}^2 + \delta_{13}q_1{q_3}^2 = 0 \\ \label{sup_dHdQ1_2}
    \pdv{H}{q_2} &= \beta_2{q_2} + \gamma_{012}p_0q_1 + \delta_{12}{q_1}^2q_2 = 0 \\
    \pdv{H}{q_3} &= \beta_3{q_3} + \gamma_{134}q_1q_4 + \delta_{13}{q_1}^2q_3 = 0 \\
    \pdv{H}{q_4} &= \beta_4{q_4} + \gamma_{134}q_1q_3 + \delta_4{q_4}^3 = 0
\end{align}
Second, optimize $q_i$ ($i\neq1$).
\begin{align}
    q_2 &= -\frac{\gamma_{012}p_0q_1}{\beta_2 + \delta_{12}{q_1}^2} \label{sup_Q2} \\
    q_3 &= -\frac{\gamma_{134}q_1q_4}{\beta_3 + \delta_{13}{q_1}^2} = -\frac{\gamma_{134}q_1\sqrt{-\frac{\beta_4}{\delta_4}}}{\beta_3 + \delta_{13}{q_1}^2} \label{sup_Q3}\\
    |{q_4}| &= \sqrt{-\frac{\beta_4}{\delta_4}}
\end{align}
Note that $q_4$ can be optimized for pre-trigger dielectric phase, as we are searching for $p_{0,c}$. Third, insert $q_i$ back to $(\partial^2 H/\partial {q_1}^2)$ and determine the condition it is negative at $q_1=0$.
\begin{gather}
   \left. \pdv[2]{H}{{q_1}} \right|_{q_1=0} = \frac{\beta_1{\beta_2}^2{\beta_3}^2 - \beta_2{\beta_3}^2{\gamma_{012}}^2{p_0}^2 - {\beta_2}^2\beta_3{\gamma_{134}}^2{q_4}^2}{{\beta_2}^2{\beta_3}^2} < 0 \\
   {p_0}^2 > \frac{\beta_1{\beta_2}{\beta_3} - {\beta_2}{\gamma_{134}}^2{q_4}^2}{{\beta_3}{\gamma_{012}}^2} = \frac{\beta_1{\beta_2} - \frac{{\beta_2}(-\beta_4){\gamma_{134}}^2}{\beta_3\delta_4}}{{\gamma_{012}}^2} \\
   |p_{0,c}| = \frac{
    \sqrt{
    \beta_1{\beta_2} - \frac{{\beta_2}(-\beta_4){\gamma_{134}}^2}{\beta_3\delta_4}
    }
    }{|\gamma_{012}|}
\end{gather}
The numerator is reduced from $\sqrt{\beta_1{\beta_2}}$ to $\sqrt{\beta_1{\beta_2} - \frac{{\beta_2}(-\beta_4){\gamma_{134}}^2}{\beta_3\delta_4}}$, so multiple coupling reduces the value of shared trigger. From Equation~(\ref{sup_Q2}) and (\ref{sup_Q3}), simultaneous condensation of $q_1$, $q_2$ and $q_3$ is shown.

What we show is that with the additional trilinear coupling, instead of an additional trigger being created leading to second phase transition, the critical value $p_{\mathrm{0,c}}$ is decreased substantially and shared among both couplings---the simultaneous condensation of $q_1$, $q_2$, and $q_3$ occurs. In other words, the hybrid mode $q_1q_2q_3$ instability occurs at a lower $p_0$ compared to just $q_1q_2$ or $q_1q_3$. 

\subsection*{Derivation of hybrid-triggered instability with polar hybrid mode}
\begin{equation} \label{sup_H2}
    H = \frac{\beta_0}{2}{p_0}^2 + \frac{\beta_1}{2}{q_1}^2 + \frac{\beta_2}{2}{q_2}^2 + \gamma p_0q_1q_2 + \frac{\delta_0}{4}{p_0}^4 + \frac{\delta_{12}}{2}{q_1}^2{q_2}^2 - v(\lambda p_0+\mu q_1q_2)
\end{equation}
First, determine the partial derivatives.
\begin{align}
    \pdv{H}{p_0} &= \beta_0{p_0} + \gamma q_1q_2 + \delta_0{p_0}^3 - \lambda v = 0 \label{sup_dHdP0_3} \\ 
    \pdv{H}{q_1} &= \beta_1{q_1} + \gamma p_0q_2 + \delta_{12}q_1{q_2}^2 - \mu vq_2 = 0 \\
    \pdv{H}{q_2} &= \beta_2{q_2} + \gamma p_0q_1 + \delta_{12}{q_1}^2q_2 - \mu vq_1 = 0
\end{align}
Second, optimize $q_i$ ($i\neq1$).
\begin{align}
    q_2 = -\frac{(\gamma p_0-\mu v)q_1}{\beta_2 + \delta_{12}{q_1}^2}
\end{align}
Here, $v$ also has to be represented as a function of $p_0$ and the coefficients. From Equation~(\ref{sup_dHdP0_3}):
\begin{gather}
    v = \frac{\beta_0{p_0}  + \delta_0{p_0}^3}{\lambda} \\
    q_2 = -\frac{[(\gamma  - \frac{\mu \beta_0}{\lambda} )p_0 - \frac{\mu\delta_0}{\lambda}{p_0}^3]q_1}{\beta_2 + \delta_{12}{q_1}^2}
\end{gather}
Note that $v$ was optimized for pre-trigger dielectric phase. Third, insert $q_i$ back to $\partial^2 H/\partial {q_1}^2$ and determine the condition it is negative at $q_1=0$. The solution is equivalent to the derivation of eq.~\ref{simple}, where $\gamma p_0$ is substituted to $\gamma p_0-\mu v$.
\begin{gather}
    \left. \pdv[2]{H}{{q_1}} \right|_{q_1=0} = 
    \frac{\beta_1{\beta_2}^2 - \beta_2(\gamma p_0-\mu v)^2}{{\beta_2}^2} < 0 \\
    \left( (\gamma  - \frac{\mu\beta_0}{\lambda} )p_0 - \frac{\mu\delta_0}{\lambda}{p_0}^3 \right)^2 = \beta_1{\beta_2}
\end{gather}
While the general solution cannot be expressed, when $\delta_0\sim0$:
\begin{gather}
    |p_{0,c}| = \frac{\sqrt{\beta_1\beta_2}}{\left| \gamma -\frac{\mu \beta_0}{\lambda} \right|}
\end{gather}
Depending on the sign of $\mu$, the polar hybrid mode can either increase or decrease $|p_{\mathrm{0,c}}|$. In the case of HfO$_2$, the direction of polarization from second order hybrid modes always oppose the direction of polarization from the polar mode. This would correspond to the same sign of $\gamma $ and $\mu$, which increases $|p_{\mathrm{0,c}}|$.

\newpage


\begin{figure} 
    \centering
    \includegraphics[width=1.0\textwidth]{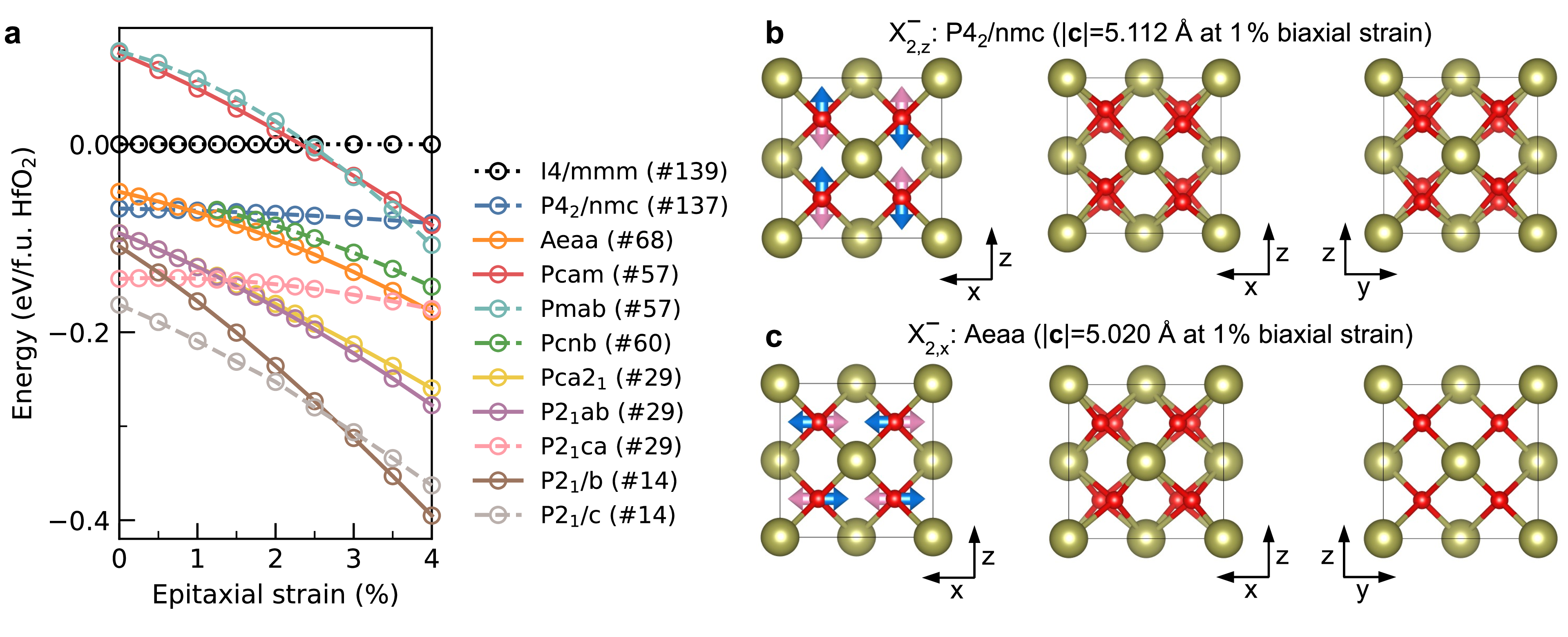} 
    \caption{\textbf{Biaxial strain phase diagram of HfO$_2$.}
    \textbf{a,} Solid lines indicate phases derived from $\mathrm{X}_{2,x}^-$, while dashed lines represent phases derived from $\mathrm{X}_{2,z}^-$.
    \textbf{b,} Structural representation of $\mathrm{X}_{2,z}^-$ mode and \textbf{c,} $\mathrm{X}_{2,x}^-$ mode. Blue arrows represent oxygen displacements in the front half of the unit cell ($y=0.75$ in direct coordinates), and the pink arrows represent those in the rear half ($y=0.25$).    
    In the absence of a monoclinic lattice transformation (as well as Pbca or domain-separated structures from ref.~\autocite{lee2020scale}), the Aeaa phase emerges as the most stable nonpolar ground state under moderate tensile strain, derived from structural instability in the fluorite HfO$_2$. 
    Experimental observations of the lattice constants from \autocite{cheema2020enhanced,cheema2022emergent} further support this. Transition from Pca2$_1$ to P4$_2$/nmc increases the interplanar spacing, whereas transition from Pca2$_1$ to Aeaa reduces it. ($c=5.038$ Å for Pca2$_1$ phase)
    }
    \label{sup_strain}
\end{figure}

\begin{figure}
    \centering
    \includegraphics[width=0.55\textwidth]{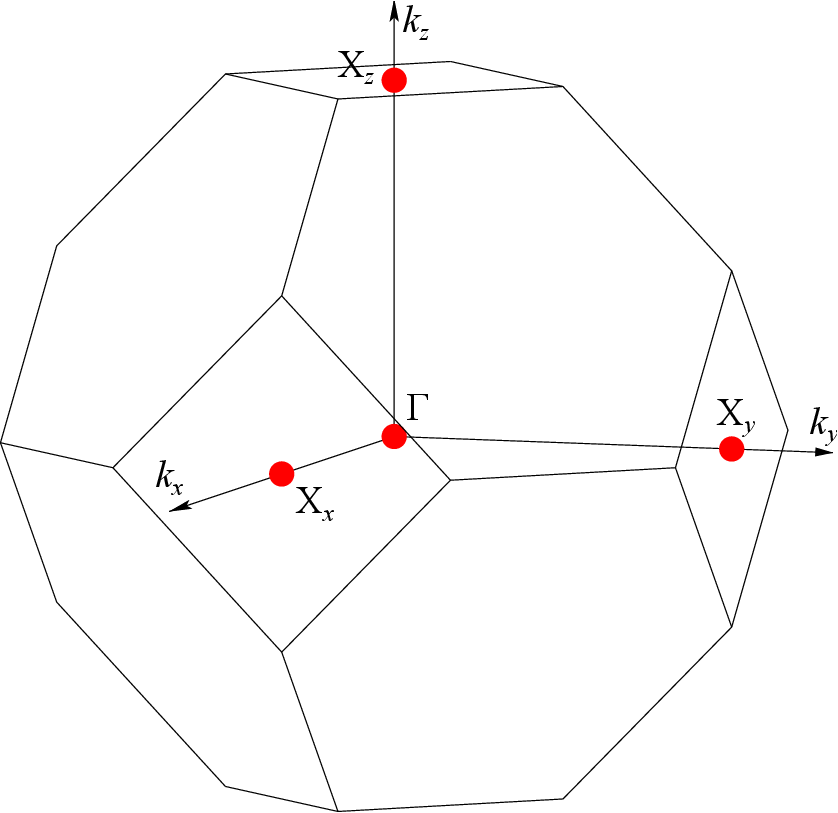} 
    \caption{\textbf{First Brillouin zone of a face-centered cubic lattice system \autocite{aroyo2011crystallography}.} 
    The coordinates of the wavevectors $\bm{k}_{\mathrm{X}_x}$, $\bm{k}_{\mathrm{X}_y}$, and $\bm{k}_{\mathrm{X}_z}$ corresponding to the positions of the $\mathrm{X}_x$, $\mathrm{X}_y$, and $\mathrm{X}_z$ points in reciprocal space, are $(2\pi/a, 0, 0)$, $(0, 2\pi/a, 0)$, and $(0, 0, 2\pi/a)$ respectively in $(k_x, k_y, k_z)$. The sum $\bm{k}_{\mathrm{X}_x} + \bm{k}_{\mathrm{X}_y} + \bm{k}_{\mathrm{X}_z}$ forms a $\Gamma$-point modulo in reciprocal space, enabling a trilinear coupling with irrep equal to that of the energy ($\Gamma_1^+$). Note that the sum of wavevectors being equal to a $\Gamma$-point modulo is a necessary condition for irrep equivalence with the energy, and not a sufficient one. For instance, the wavevector sum arising from the bilinear coupling $Q_1Q_5$ ($\bm{k}_{\mathrm{X}_x} + \bm{k}_{\mathrm{X}_x}$) also forms a $\Gamma$-point modulo but exhibits irrep equivalent to polarization ($\Gamma_4^-$). Consequently, the bilinear coupling does not directly appear in the energy expression.} 
    \label{sup_BZ}
\end{figure}

\begin{figure}
    \centering
    \includegraphics[width=0.55\textwidth]{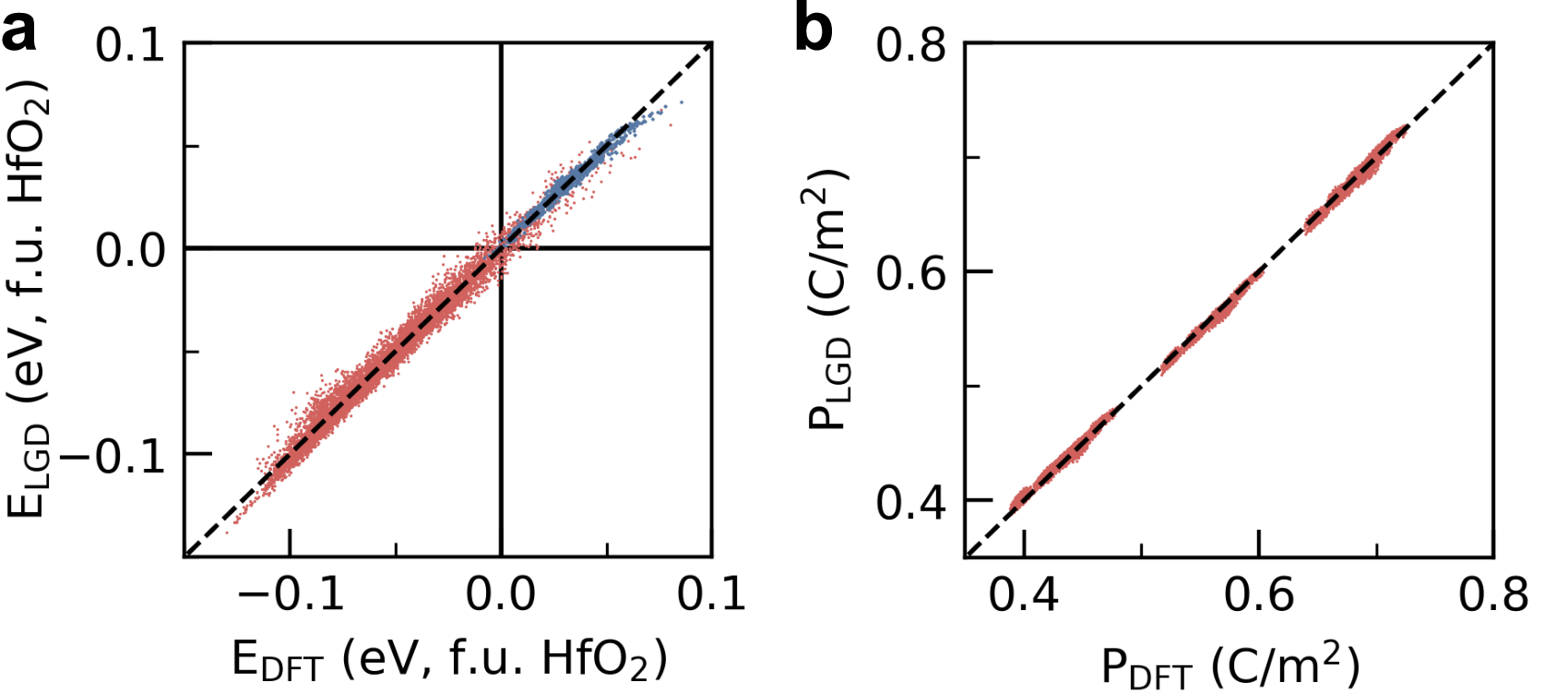} 
    \caption{\textbf{Symmetry-adapted regression of DFT data using LGD theory}
    \textbf{a,} Energy and \textbf{b,} polarization data of HfO$_2$ under 1\% tensile biaxial strain. Blue points correspond to 3$^8$ data points around the high-symmetry I4/mmm phase, while red points correspond to 3$^8$ data points around the ferroelectric Pca2$_1$ phase.
    } 
    \label{sup_reg}
\end{figure}

\begin{figure}
    \centering
    \includegraphics[width=0.65\textwidth]{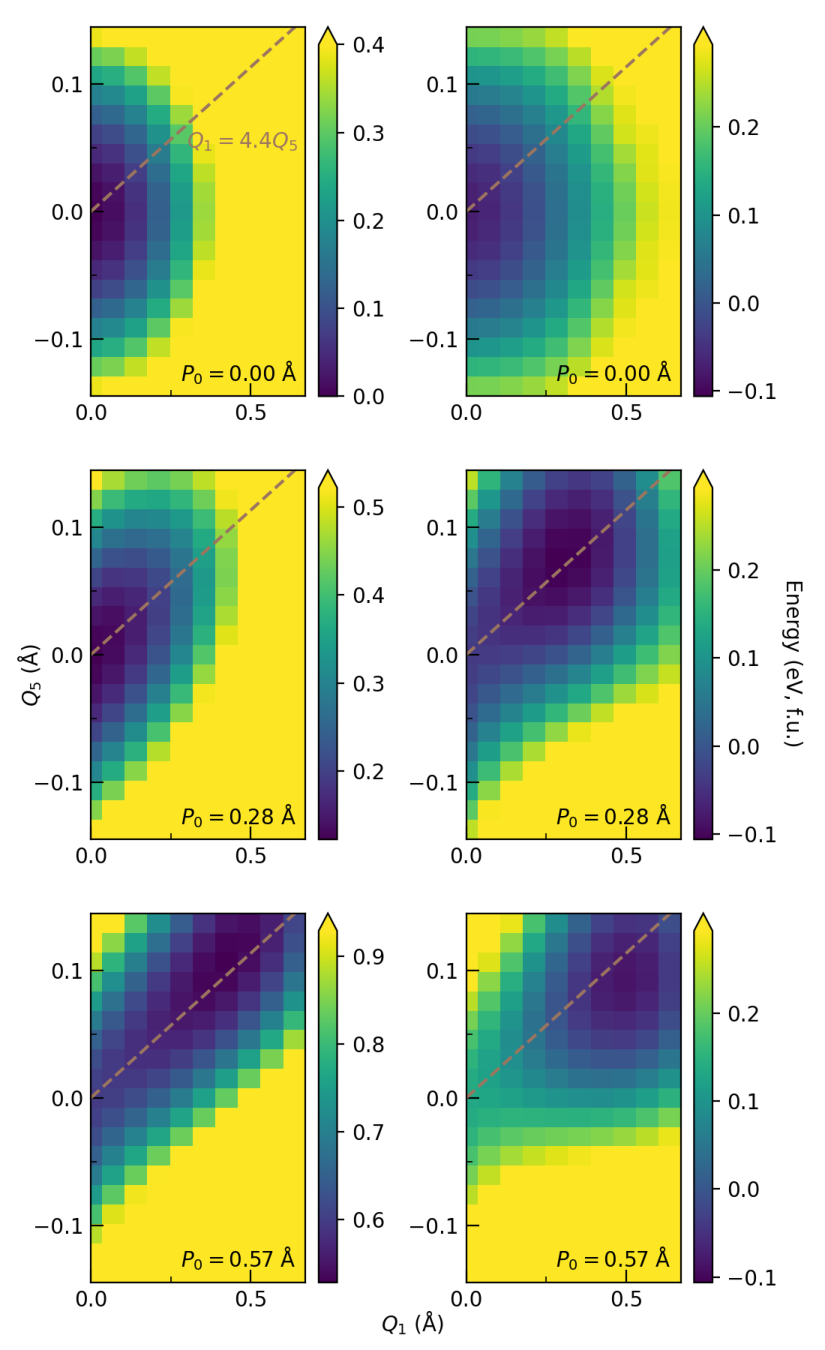} 
    \caption{\textbf{Two-dimensional DFT-calculated energies of HfO$_2$, illustrating hybrid-triggered ferroelectricity and $P_0Q_1Q_5$ coupling.} 
    The results compare scenarios where all other order parameters are fixed to zero (left) and fully relaxed (right). One-dimensional data shown in Fig.~\ref{multicoupling}c is along $Q_1=4.4Q_5$ line which preserve the ratio at the ferroelectric minimum. The fixed ratio is not necessarily imposed from the hybrid-triggered ferroelectricity. Note that even when the hybrid mode $Q_1Q_5$ instability appears, individual modes $Q_1$ and $Q_5$ can remain hard.
    } \label{sup_015_2d}
\end{figure}

\begin{figure}
    \centering
    \includegraphics[width=0.3\textwidth]{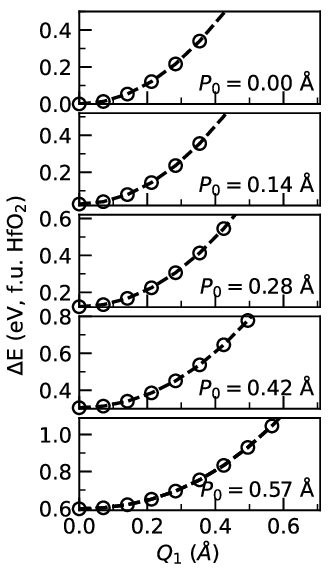} 
    \caption{\textbf{Biquadratic coupling between $P_0$ and $Q_1$.}
    A slight negative biquadratic coupling exists between $P_0$ and $Q_1$ \autocite{zhou2022strain}. However, in contrast to the trilinear and quadlinear couplings shown in Fig. \ref{multicoupling}, this coupling between the two hard modes does not induce a new local minimum or phase transition. The negative biquadratic coupling also occurs for ${Q_1}^2{Q_4}^2, {Q_3}^2{Q_4}^2, {P_0}^2{Q_5}^2, {Q_3}^2{Q_5}^2, {Q_1}^2{Q_6}^2, {Q_1}^2{Q_7}^2$, and ${Q_4}^2{Q_7}^2$ without major contribution to energy.}  \label{sup_01coupling}
\end{figure}

\begin{figure}
    \centering
    \includegraphics[width=0.9\textwidth]{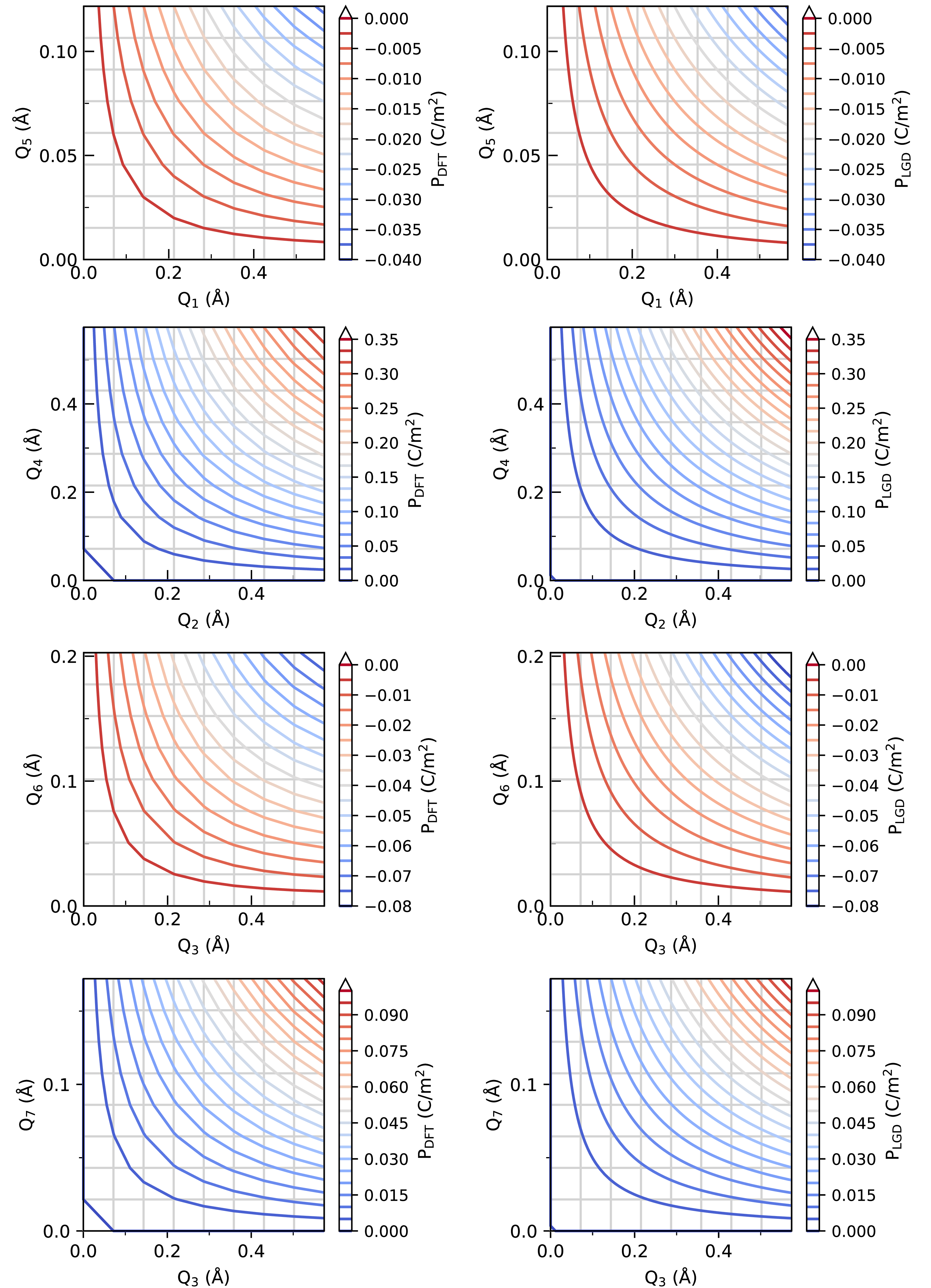} 
    \caption{\textbf{Polarization from hybrid nonpolar modes.} 
    Polarization of 1\% biaxial strained HfO$_2$ from DFT calculations (left) and symmetry-guided LGD theory up to second order (right).
    } \label{sup_charge_regression}
\end{figure} 

\begin{figure}
    \centering
    \includegraphics[width=0.5\textwidth]{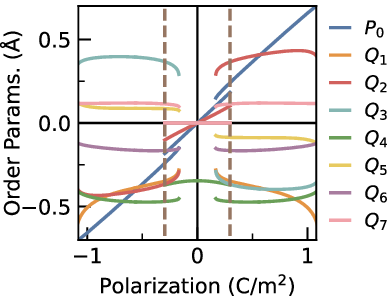} 
    \caption{\textbf{Alternative domain switching of HfO$_2$.} 
    Despite the presence of 8 modes, each triggered phase only have 2 domains that share the same sign of $Q_4$. The domains not shown in Fig. \ref{charge_effect}B is displayed here. There is equal chance of the modes condensing to each domain when the ferroelectricity is triggered.
    } \label{sup_domain}
\end{figure}

\clearpage

\begin{table}[!]
	\centering
	\caption{\textbf{Irreducible representation (irrep) of order parameters from different parent structures.}
	Hf ion at (0, 0, 0) and O ion at (0.25, 0.25, 0.25) in direct coordinates from high-symmetry structure is used as reference for in-phase Hf and O. Lattice transformation from Fm$\bar{3}$m to I4/mmm requires rotation of Cartesian bases around $z$ axis by $\pi/4$.}
	\label{sup_table} 

    \renewcommand{\arraystretch}{1.5}
	\begin{tabular}{cccccccc} 
		\\
		\hline
		  \multirow{2}{*}{Order Parameter} & \multicolumn{2}{c}{Irrep} & \multicolumn{4}{c}{Ionic displacements} & \multirow{2}{*}{Space Group} \\
             & Fm$\bar{3}$m & I4/mmm & Hf$_{\mathrm{in}}$ & 
  Hf$_{\mathrm{out}}$ & O$_{\mathrm{in}}$ & O$_{\mathrm{out}}$ \\
		\hline
            $\eta_{zz}$ & $\Gamma_{1}^+ \oplus \Gamma_{3}^+$ & $\Gamma_{1}^+$ & - & - & - & -  & I4/mmm (\#139) \\
		$P_0$ & $\Gamma_{4}^-$ & $\Gamma_{3}^-$ & - & - & $-u_z$ & - & I4mm (\#107)\\
		$Q_1$ & $\mathrm{X}_{5,y}^+$ & $\mathrm{X}_{3,y}^+$  & - & - & $u_z$ & $-u_z$ & Bmeb (\#64)\\
		$Q_2$ & $\mathrm{X}_{5,x}^+$ & $\mathrm{X}_{2,x}^+$ & - & - & $u_y$ & $-u_y$ & Aeam (\#64)\\
		$Q_3$ & $\mathrm{X}_{5,z}^+$ & $\mathrm{M}_{5}^+$ & - & - & $u_x$ & $-u_x$ & Ccme (\#64)\\
		$Q_4$ & $\mathrm{X}_{2,x}^-$ & $\mathrm{X}_{1,x}^-$ & - & - & $u_x$ & $-u_x$ & Aeaa (\#68)\\
		$Q_5$ & $\mathrm{X}_{3,y}^-$ & $\mathrm{X}_{4,y}^-$ & $u_y$ & $-u_y$ & - & - & Bmem (\#67) \\
		$Q_6$ & $\mathrm{X}_{5,z}^-$ & $\mathrm{M}_{5}^-$ & $u_x$ & $-u_x$ & - & - & Ccmm (\#63)\\
		$Q_7$ & $\mathrm{X}_{5,z}^-$ & $\mathrm{M}_{5}^-$ & - & - & $u_y$ & $-u_y$ & Ccmm (\#63)\\
		\hline
	\end{tabular}
\end{table}

\begin{table}[!]
    \centering
    \caption{\textbf{Inversion breaking of the symmetry-modes.} $i_1$ to $i_4$ denote the inversion centers present in the high-symmetry conventional fluorite unit cell. For every inversion center $(x,y,z)$, there are 7 more translation induced inversion centers at $(x+0.5,y,z),(x,y+0.5,z),(x,y,z+0.5),(x,y+0.5,z+0.5),(x+0.5,y,z+0.5),(x+0.5,y+0.5,z),$ and $(x+0.5,y+0.5,z+0.5)$. Symbol $\bigcirc$ indicates that the inversion center is preserved upon the condensation of the mode, while $\times$ signifies that the inversion center is destroyed.}
    \label{sup_inversion} 

    \renewcommand{\arraystretch}{1.5}
	\begin{tabular}{ccccccccccccc} 
		\\
		\hline
		  \multirow{2}{*}{Inversion center} & \multicolumn{3}{c}{Direct coordinates} & \multirow{2}{*}{$\eta_{zz}$} & \multirow{2}{*}{$P_0$} & \multirow{2}{*}{$Q_1$} & \multirow{2}{*}{$Q_2$} & \multirow{2}{*}{$Q_3$} & \multirow{2}{*}{$Q_4$} & \multirow{2}{*}{$Q_5$} & \multirow{2}{*}{$Q_6$} & \multirow{2}{*}{$Q_7$} \\
             & $x$ & $y$ & $z$ & & & & & & & &  \\
		\hline
		$i_1$ & 0 & 0  & 0      & $\bigcirc$& $\times$ & $\bigcirc$ &$\bigcirc$&$\bigcirc$& $\times$ & $\times$ & $\times$ & $\times$ \\
		$i_2$ & 0 & 0.25 & 0.25 & $\bigcirc$& $\times$ & $\times$ &$\bigcirc$& $\times$ & $\times$ &$\bigcirc$&$\bigcirc$&$\bigcirc$\\
		$i_3$ & 0.25 & 0 & 0.25 & $\bigcirc$& $\times$ &$\bigcirc$& $\times$ & $\times$ &$\bigcirc$& $\times$ &$\bigcirc$&$\bigcirc$\\
		$i_4$ & 0.25 & 0.25 & 0 & $\bigcirc$& $\times$ & $\times$ & $\times$ &$\bigcirc$&$\bigcirc$&$\bigcirc$& $\times$ & $\times$ \\
		\hline
	\end{tabular}
\end{table}

\printbibliography
\end{refsection}

\end{document}